\documentclass[aps,prb, superscriptaddress, notitlepage,twocolumn,10pt,nobalancelastpage]{revtex4-2}
\usepackage{graphicx} 
\usepackage{hyperref}
\usepackage{caption}
\usepackage{subcaption}
\usepackage{bm}
\hypersetup{ 
    colorlinks=true, 
    linktoc=all,     
    linkcolor=blue,  
    citecolor =blue
} 
\usepackage{amsmath}
\usepackage{upgreek}

\DeclareMathOperator{\Tr}{Tr}
 
\renewcommand{\v}[1]{\bm{#1}} 
\newsavebox{\imagebox}

\begin{document}

\title{Interplay of interactions and disorder at the charge-density wave  transition of two-dimensional Dirac semimetals}

\author{Mikolaj D. Uryszek}
\affiliation{London Centre for Nanotechnology, University College London, Gordon St., London, WC1H 0AH, United Kingdom}
\author{Frank Kr\"uger}
\affiliation{London Centre for Nanotechnology, University College London, Gordon St., London, WC1H 0AH, United Kingdom}
\affiliation{ISIS Facility, Rutherford Appleton Laboratory, Chilton, Didcot, Oxfordshire OX11 0QX, United Kingdom}

\begin{abstract}
We consider the effects of weak quenched fermionic disorder on the quantum-phase transition between the Dirac semimetal and charge density wave (CDW) insulator in 
two spatial dimensions. The symmetry breaking transition is  described by the Gross-Neveu-Yukawa (GNY) theory of Dirac fermions coupled to an Ising order parameter 
field.  Treating the disorder using the replica method, we consider chemical potential,  vector potential (gauge), and random mass disorders, which all arise from non-magnetic 
charge impurities. We self-consistently account for the Landau damping of long-wavelength order-parameter fluctuations by using the non-perturbative RPA re-summation of 
fermion loops, and compute the renormalization-group (RG) flow to leading order in the disorder strength and $1/N$ ($N$ the number of Dirac fermion flavors). 
We find two fixed points, the clean GNY critical point which is stable against weak disorder, and a dirty GNY multi-critical point, at which the chemical potential disorder is finite 
and the other forms of disorder are irrelevant. We investigate the scaling of physical observables at this finite-disorder multi-critical point which breaks Lorentz 
invariance and gives rise to distinct non-Fermi liquid behavior.   
\end{abstract}

\maketitle
\section{Introduction} 

Over the past couple of decades, two and three dimensional topological semimetals have been at the forefront of research in condensed matter physics \cite{hasan_kane_top_insulator_review,vafek_vishwanath_diracF_in_materials_review,armitage_review_semimetals}. Within this family of systems, Dirac fermions have been found to be ubiquitous, where the most famous example of a host system is graphene - a two dimensional carbon monolayer which exhibits a four-fold degenerate band crossing at charge neutrality.  Due to the unique properties of Dirac/Weyl semimetals, e.g. point-like Fermi surface and linearly vanishing density of states, they are well suited to realizations of high energy phenomena in an accessible condensed matter setting.

In any realistic condensed matter system disorder is present, hence its understanding is paramount. Quenched, non-dynamical disorder has been widely studied in the non-interacting limit of
systems that  exhibit two-dimensional Dirac fermions, e.g. degenerate (or zero-gap) semiconductors \cite{fradkin_semiconductor_disorder_I,fradkin_semiconductor_disorder}, graphene \cite{efetov_transport_disorder_graphene,mirlin_transport_disordered_graphene,mirlin_minimial_conductivity_quantum_critical_graphene,das_sarma_carrier_transport_graphene,
das_sarma_graphene_transport_mean_field_theory,aleiner_graphene_conductivity,rossi_transport_graphene_review}, and $d$-wave superconductors \cite{wenger_dos_disorder_2d_Dirac,wenger_disorder_defetcs_in_d-wave_SC,altland_d_wave_disorder}. 

A lot of interest was triggered by the first graphene experiments \cite{Novoselov+04,Novoselov+05,Zhang+05} which showed a minimal conductivity of the order of the conductance 
quantum $e^2/h$ over a wide range of 
temperatures. It was shown theoretically  that the transport properties depend crucially on the type of disorder \cite{mirlin_transport_disordered_graphene} but that for randomness which preserves 
one of the chiral symmetries of the clean Hamiltonian the conductivity is equal to the minimal value \cite{mirlin_transport_disordered_graphene,mirlin_minimial_conductivity_quantum_critical_graphene}, suggesting that the transport is not affected by localization and 
remains ballistic. However, this universal result is based on a self-consistent Born approximation, which is not applicable to massless Dirac fermions in two spatial dimensions \cite{efetov_transport_disorder_graphene,wenger_dos_disorder_2d_Dirac,wenger_disorder_defetcs_in_d-wave_SC}. More recently, it was argued that over the experimentally accessible temperature
range, graphene is in the Drude–Boltzmann diffusive transport regime and that density inhomogeneities from remote charge impurities render the Dirac points effectively inaccessible to 
experiments \cite{das_sarma_carrier_transport_graphene,das_sarma_graphene_transport_mean_field_theory}. 
Using a self consistent RPA-Boltzmann approach, the authors showed that the conductivity is indeed of order $e^2/h$ but with a non-universal pre-factor that depends on the disorder 
distribution. Remote charge impurities can be viewed as random chemical potential shifts that give rise to puddles of electron and hole-doped regions in the graphene layer. 
Building on that picture, the scaling of the conductivity was obtained within a random resistor network model that describes the percolation of $p$- and $n$-type regions \cite{aleiner_graphene_conductivity}. 

As the minimal conductivity puzzle shows, there is a lot of rich physics already at the non-interacting level. However, an accurate description of a Dirac semimetal also must include the effects of 
electron-electron interactions, on top of the disorder. 
For weak Coulomb interactions the clean two-dimensional Dirac fixed point is unstable against generic disorder and the RG flow is dominated by the randomness in the chemical potential \cite{sachdev_coulomb_quantum_hall,ye_weak_disorder_quantum_hall,vozmediano_disorder_graphene}, similar to the non-interacting case \cite{efetov_transport_disorder_graphene} and 
consistent with the picture of local electron and hole ``puddles''. On the other hand, in the regime of moderate to strong Coulomb interactions, it was found that fluctuations associated with such random 
potential disorder are parametrically cut off by screening and that instead the runaway flow is dominated by vector potential disorder \cite{aleiner_graphene_large_N}. Such disorder 
 from elastic lattice deformations (``ripples") \cite{mirlin_transport_disordered_graphene} and topological lattice defects \cite{Gonzalez+93,Lammert+00,Gonzalez+01}. 

The situation is very different in three-dimensional Dirac/Weyl semimetals with long-range Coulomb interactions. In these systems the semi-metallic phase is stable against short-range correlated disorder. Above
a critical disorder strength, the semi-metallic phase undergoes a quantum phase transition into a disorder controlled diffusive metallic phase with a finite density of states at the Fermi level \cite{chakravarty_diffusive_metal,gonzalez_3d_weyl_coulomb_disorder,das_sarma_disorder_3d_dirac,wang_weyl_3d_coulomb_disorder,Syzranov+18}. 
It remains a controversial issue whether the disorder transition is rounded out by non-perturbative, rare region effects \cite{sondhi_rare_region_effects_dirac_3D,pixley_avoided_QCP_weyl,Pixley+16} 
or not \cite{altland_dos_weyl_rare_region,altland_weyl_stable_against_disorder}.

Under a sufficiently strong short-ranged electron-electron interaction a Dirac semimetal will undergo a quantum phase transition into a symmetry broken state where the fermionic spectrum is gapped. 
Such a transition is best described using a composite fermion-boson approach, resulting from a Hubbard-Stratonovich decoupling of the fermionic interaction vertex in the 
relevant channel with a dynamical order parameter field. In the case of Dirac fermions the resulting field theory is known as the Gross-Neveu-Yukawa (GNY) model which describes chiral symmetry 
breaking and spontaneous mass generation in high energy physics \cite{gross_neveu_1974,zinn-justin_four_fermion_interaction_1991}. The symmetry broken phase is dependent on the nature of the microscopic interactions; for the half-filled Hubbard model on the honeycomb lattice with competing interactions a vast array of phases were found \cite{grushin_graphene_phases}, including 
antiferromagnetism, different types of charge order, Kekule phases and topological Quantum Hall states. 
 
The effects of weak quenched disorder on the semimetal-to-superconductor transition, described by the XY GNY model, were studied using $\epsilon$ expansions below the 
upper critical dimension \cite{sondhi_sc_of_disordered_Dirac,maciejko_dirac_disorder_double_epsilon}. It was found that chemical potential disorder is strongly irrelevant at the clean 
quantum-critical point in $D=4-\epsilon$ space-time dimensions but that disorder in the superconducting order parameter mass plays a crucial role. Such bosonic disorder would arise 
from randomness in the attractive fermion interaction after Hubbard-Stratonovich decoupling. In the supersymmetric case of a single two-component Dirac field coupled to the XY order parameter, 
there is a marginal flow away from the clean critical point to strong disorder \cite{sondhi_sc_of_disordered_Dirac,maciejko_dirac_disorder_double_epsilon}.
However, if degeneracies such as spin or valley pseudo-spin are included, the clean fixed point becomes stable against weak bosonic mass disorder and a finite-disorder multi-critical point 
with non-integer dynamical exponent ($z>1$) can be identified within the double $\epsilon$ expansion \cite{maciejko_dirac_disorder_double_epsilon}. 
Similar finite disorder fixed points were established in the chiral Ising and Heisenberg GNY models with bosonic random-mass disorder, using triple $\epsilon$ expansion \cite{maciejko_gny_random_mass_disorder}. 

In this work we revisit the effects of disorder on the quantum criticality of two dimensional Dirac/Weyl fermions. For simplicity, we focus on quantum phase transitions that, in the absence of disorder, are described by the chiral Ising GNY theory. An example is the CDW transition of electrons on the half-filled honeycomb lattice that is driven by a repulsive nearest-neighbor interaction and characterized by an imbalance of charge on the two sublattices. Our work departs in two important aspects from previous studies \cite{sondhi_sc_of_disordered_Dirac,maciejko_dirac_disorder_double_epsilon,maciejko_gny_random_mass_disorder}. Firstly, we omit an $\epsilon$ expansion and compute the quantum corrections 
in two spatial dimensions. Away from the upper critical dimension, the Landau damping of long-wavelength order-parameter fluctuations is a non-perturbative effect. It renders the order parameter propagator non-analytic in the IR limit, thereby changing the universal critical behavior \cite{uryszek_soft_cut_off,chubukov_landau_damping}. This physics is not captured by the $\epsilon$ expansion 
since the boson propagator remains analytic at the upper critical dimension. Secondly, we consider disorder on the level of the original fermionic theory, e.g. in the form of a random potential from point-like impurities. In the physical dimension, such fermionic potential disorder is marginal at the clean GNY fixed point.  

Our paper is organized as follows. In Sec.~\ref{sec:model} we present the Yukawa theory for Dirac fermions subject to a strong-short ranged interaction and uncorrelated disorder. We utilize the 
RPA to account for the non-perturbative Landau damping of long-wavelength order-parameter fluctuations. The disorder is treated using the replica formalism \cite{edwards_anderson_spin_glasses_1975,fischer_hertz_sping_glasses}. In Sec.~\ref{sec:RG_analysis} we employ Wilson's momentum shell RG 
within the large-$N$ expansion to derive the flow of the disorder couplings, and compute the critical exponents to leading order in $1/N$ and weak disorder. Lastly, in Sec.~\ref{sec:discussion} we 
present a summary of our main findings, and compare them with previous literature.

\section{Model}\label{sec:model}
\subsection{Clean Dirac fermions in 2+1 dimensions}
We consider Dirac Fermions with dispersion $\epsilon(\v{k}) = \pm v |\v{k}|$ in two spatial dimensions, described by the imaginary time action 
\begin{equation}\label{eqn:dirac_noninteracting}
S_\psi = \int d^{2}\v{x}\int d\tau \,  \Psi^{\dag} \left(\partial_\tau+ i v \v{\partial}\cdot\bm{\sigma}\right)\Psi,\\
\end{equation}
over fermionic Grassmann fields $\Psi(\v{x},\tau)$. Here  $\v{\partial}=(\partial_x,\partial_y)$ and  $\v{\sigma}=(\sigma^x,\sigma^y)$ are the conventional 
$2\times2$ Pauli matrices. This action describes non-interacting 
electrons on the half-filled  honeycomb lattice in the long-wavelength, low-energy limit, where in this case the Pauli matrices act on the $\{A,B\}$ sublattice pseudospin
subspace. In addition, the fermionic Grassmann fields carry the electron spin flavors and the valley indices from the two distinct Dirac points in the Brillouin zone.  

In the following, we do not consider spontaneous symmetry breaking or disorder that lift the spin and valley degeneracies. We further generalize to a total number of
$N$ components of the fermion fields, $ \Psi = (\psi_1,\ldots,\psi_N)$,  in order to gain analytic control through an expansion in $1/N$. For brevity, we use the short-hand notation 
$\Tr[\sigma^i \sigma^j] = N\delta_{ij}$.

We consider the case where strong short-range  interactions drive an instability in the charge channel, which corresponds to a quantum phase transition from a Dirac semimetal to a CDW insulator where the sublattice symmetry is spontaneously broken. Generally this transition belongs to the chiral Ising GNY universality class \cite{gross_neveu_1974,zinn-justin_four_fermion_interaction_1991}, which is best studied within the Yukawa language where the Dirac fermions couple to a real-valued, scalar dynamical order parameter 
$\phi(\v{x},\tau)$. This results in the chiral Ising GNY model, 
\begin{equation}
S_{\text{GNY}} = S_\psi + S_g + S_\phi + S_\lambda,
\end{equation}
where 
\begin{align}
S_g &= \frac{g}{\sqrt{N}}\int d^2\v{x}\int d\tau\ \phi \Psi^\dagger \sigma^z \Psi,\\
S_\phi &= \frac{1}{2}\int d^2\v{x}\int d\tau\ \phi(-\partial_\tau^2 - c^2 \v{\partial}^2 + m^2) \phi,\label{eqn:boson_non_interacting_naive_action}\\
S_\lambda &= \frac{\lambda}{N}\int d^2\v{x} \int d\tau\ \phi^4.
\label{eq.phi4}
\end{align}

Starting from a lattice model on the honeycomb lattice, the Yukawa coupling $S_g$ arises naturally from a Hubbard Stratonovich decoupling of a repulsive nearest-neighbor 
interaction. The Yukawa coupling anti-commutes with the non-interacting action, Eq.(\ref{eqn:dirac_noninteracting}), and thereby fully gaps the fermionic spectrum in the ordered 
phase where $\langle\phi\rangle\neq0$. Here $c$ is the bosonic velocity, and $m^2$ is the tuning parameter for the quantum phase transition. In the context of the CDW transition on the 
honeycomb lattice, $m^2\sim V_c - V$, where $V$ is the repulsion between electrons on adjacent sites and $V_c$ the critical interaction strength.

\subsection{Landau damping of order-parameter fluctuations in $d=2$}

\begin{figure}[t]
\centering
\includegraphics[width=0.8\columnwidth]{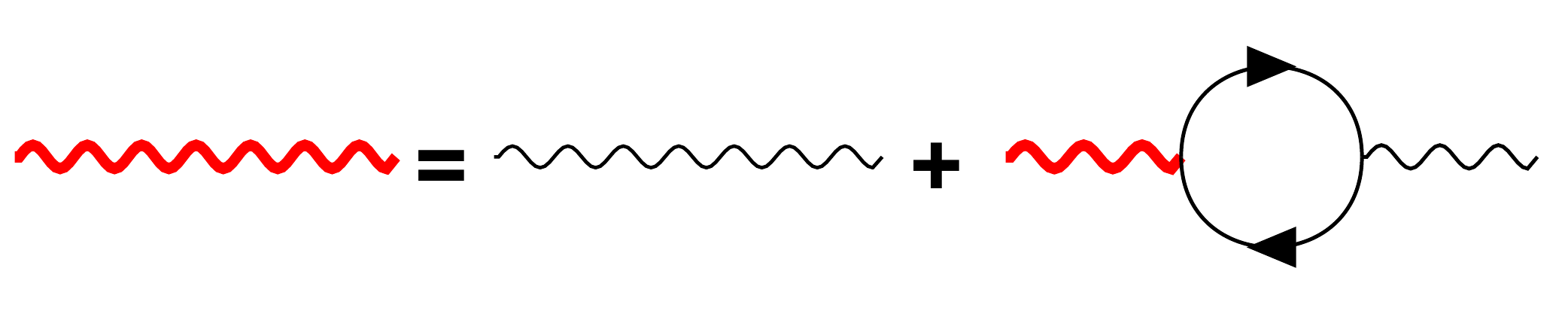}
\caption{Random phase approximation to account for the non-perturbative Landau damping of IR order-parameter fluctuations. The black wavy line represents the bare order parameter, while the fermionic propagator is denoted by the solid arrowed line.}
\label{fig:RPA}
\end{figure}

The functional form of Eq.~(\ref{eqn:boson_non_interacting_naive_action}) is obtained naively from considering the most relevant analytical behavior that the boson can exhibit. At the upper critical dimension, which for the GNY theory of Dirac fermions is $d=3$, this is sufficient. However when considering systems in physical dimensions like in $d=2$, as done in this work, it is imperative to consider the phenomenon of Landau damping of the order parameter fluctuations by gapless electronic particle-hole excitations. To self-consistently account for these damped dynamics, we use the non-perturbative RPA resummation of fermion loops, shown diagrammatically in Fig.~\ref{fig:RPA}, to obtain the dressed inverse boson propagator. In $d=2$ the bosonic self energy is given by,
\begin{equation}
\Pi(\vec{k}) = \frac{g^2}{N}\int \frac{d^3q}{(2\pi)^3} \Tr\left[\sigma^z G_\psi(\vec{k}+\vec{q})\sigma^z G_\psi(\vec{q})\right]
\end{equation}
where $\vec{k}=(\v{k},\omega)$ and the fermion propagator is given by
\begin{equation}\label{eqn:fermion_propagator}
G_\psi(\v{k},\omega)= \frac{i \omega + v \v{k}\cdot \v{\sigma}}{\omega^2 + v^2 \v{k}^2}.
\end{equation}

Since the leading IR behavior of the integral for $\Pi(\v{k},\omega)$ is independent of the UV momentum cut-off we can send the cut-off to infinity and perform the integration over
the entire three dimensional frequency-momentum space. This leads to the result \cite{uryszek_soft_cut_off}
\begin{equation}
\Pi(\v{k},\omega) = \frac{g^2}{16 v^2}\left(\omega^2 + v^2 \v{k}^2\right)^{1/2}. 
\end{equation}

In the long-wavelength limit of small frequency and momenta the self energy dominates over the $(\omega^2 + c^2 \v{k}^2)$ terms in the bare inverse propagator (\ref{eqn:boson_non_interacting_naive_action}).
In order to identify the universal critical behavior we drop the sub-leading terms, which are irrelevant in an RG sense,  and use the inverse bosonic propagator 
\begin{equation}\label{eqn:boson_dressed_propagator}
G_\phi^{-1}(\v{k},\omega)  = \Pi(\v{k},\omega) + m^2.
\end{equation}

Formally, the RPA contribution dominates in the large-$N$ limit, which is evident after making the rescaling $g^2 \to g^2 N$.
The Landau damped dynamics affects the scaling of the effective order parameter field. Crucially, the quartic $\phi^4$ term of Eq.~(\ref{eq.phi4}) is rendered irrelevant at tree level and so is neglected in the following. 
This is a common feature of the ``interaction driven scaling" \cite{Lee+09} of gapless fermionic systems.

\subsection{Coupling to Disorder: Replica Field Theory}
We will consider different forms of quenched disorder fields $V_i(\v{x})$ that arise from non-magnetic charge impurities and are expected to affect the quantum phase transition between the Dirac semimetal 
and CDW insulator. These fields couple to the fermions in the different channels of the $2\times2$ sublattice pseudospin space, 
\begin{equation}
S_{\text{dis}} = \sum_{i=0,x,y,z} \int d^2\v{x}\int d\tau V_i(\v{x}) \Psi^\dagger(\v{x},\tau) \sigma^i \Psi(\v{x},\tau),
\label{eq.dis}
\end{equation} 
where in addition to the three Pauli matrices we have defined the identity matrix as $\sigma^0$. Other forms of disorder which would break degeneracies of the other fermion flavors, e.g. spin or 
valley degeneracies, are not considered here. Note that for simplicity, we do not consider disorder that couples to the bosonic order parameter.

$V_0$ and $V_z$ are random potentials that couple to the symmetric ($\psi_A^\dagger\psi_A + \psi_B^\dagger\psi_B$) and 
anti-symmetric ($\psi_A^\dagger\psi_A - \psi_B^\dagger\psi_B$) combinations of the local electron densities on the two sites in the unit cell. 
The latter combination is required as some charge impurities will affect the two sites differently and locally break the symmetry between the two sub-lattices. 
In the following, we will refer to $V_0$ as ``chemical potential disorder" since it can be viewed as spatial variations of the homogeneous chemical potential $\mu=0$, and to $V_z$ 
as ``random mass disorder" since it couples in the same way as the electronic mass gap generated by the condensation of the CDW order parameter.   

The components $V_\perp:=V_x=V_y$ correspond to random gauge (vector) potential disorder. As discussed in the context of graphene, the random gauge potential describes elastic lattice deformations or ripples \cite{aleiner_graphene_large_N,vafek_coulomb_ripples_graphene,case_2d_Dirac_Coulomb_random_gauge_potential}, which will be caused by impurity atoms. 
The different disorder fields $V_i$ are present in any system with non-magnetic impurities and, as we will show later,  there exists a rich interplay between them.

We assume that the random potentials $V_i(\v{x})$ are uncorrelated and that they follow Gaussian distributions with zero mean 
and variances $\Delta_i\ge 0$, 
\begin{align}
\langle V_i(\v{x})\rangle_\textrm{dis} &= 0,\label{eq.av1}\\
\langle V_i(\v{x}_1)V_j(\v{x}_2)\rangle_\textrm{dis} &= \Delta_i\delta_{ij}\delta(\v{x}_1-\v{x}_2),\label{eq.av2}
\end{align}
where $\langle\ldots\rangle_\textrm{dis}$ denotes the average over the disorder.
The presence of disorder on the level of the quadratic fermion action, Eq.~(\ref{eq.dis}), does not affect the Hubbard-Stratonovich decoupling of the fermion interaction. The resulting field theory is therefore given by $S_{\text{GNY}}[\Psi^\dagger,\Psi,\phi]+S_{\text{dis}}[\Psi^\dagger,\Psi]$. It is important to stress that disorder does not enter in the 
bosonic sector of the theory, e.g. in the form of random-mass disorder of the CDW order parameter field $\phi$.   In order to average the free energy over the quenched disorder, 
we use the replica trick \cite{edwards_anderson_spin_glasses_1975,fischer_hertz_sping_glasses}, 
\begin{equation}
\langle F \rangle_\textrm{dis} = -T \langle \ln Z \rangle_\textrm{dis} = -T\lim_{n\to 0} \frac{\langle Z^n \rangle_\textrm{dis}-1}{n},
\end{equation}
where $Z = \int\mathcal{D}[\Psi^\dagger,\Psi,\phi] e^{-(S_{\text{GNY}}+S_{\text{dis}})}$ denotes the partition function. After taking $n$ replicas of the 
 system and performing the average over the uncorrelated Gaussian disorder, using Eqs. (\ref{eq.av1}) and (\ref{eq.av2}), we obtain the effective replica field theory
\begin{align}
S &= \sum_{a=1}^{n}\int d^{2}\v{x} \int d\tau \,  \Psi^{\dag}_a \left(\partial_\tau+ i v \v{\partial}\cdot\bm{\sigma}+ \frac{g}{\sqrt{N}}\phi_a \sigma^z\right)\Psi_a\nonumber\\
&\;+ \frac{1}{2}\sum_{a=1}^n\int_{|\v{k}|\le\Lambda} \frac{d^2\v{k}}{(2\pi)^2}\int_{-\infty}^\infty \frac{d\omega}{2\pi}\ \ G_\phi^{-1}(\v{k},\omega)|\phi_a(\v{k},\omega)|^2\nonumber\\
&\;- \frac12 \sum_{a,b=1}^n \int d^2\v{x} \int d\tau \int d\tau^\prime \sum_{i=0,x,y,z} \Delta_i \nonumber\\
&\;\times\bigg[ \Psi_{a}^{\dag}(\v{x},\tau) \sigma^i \Psi_{a}(\v{x},\tau)\bigg]  \bigg[ \Psi_{b}^{\dag}(\v{x},\tau^\prime)\sigma^i \Psi_{b}(\v{x},\tau^\prime)\bigg],
\label{eqn:full_replica_action}
\end{align}
at zero temperature. Here $G_\phi^{-1}(\omega,\v{k})$, defined in Eq.(\ref{eqn:boson_dressed_propagator}), is the inverse dressed bosonic propagator that is obtained by the RPA resummation 
as seen in Fig.~\ref{fig:RPA}. Unlike Refs.~\cite{sondhi_sc_of_disordered_Dirac,maciejko_dirac_disorder_double_epsilon,maciejko_gny_random_mass_disorder}, we do not include a four-boson disorder vertex. Such a vertex would arise from 
a replica average of random-mass disorder of the CDW order parameter field $\phi$ which is not present in our theory. In Appendix \ref{appx:bosonic_disorder_generation}, we show that starting with the 
bare replica action (\ref{eqn:full_replica_action}),  a four-boson disorder vertex is not generated under the RG at two-loop order.

\section{Renormalisation Group analysis}\label{sec:RG_analysis}

In the following we perform a momentum-shell RG analysis of the replica action (\ref{eqn:full_replica_action}). We integrate out fast modes with momenta from an infinitesimal 
shell $\Lambda e^{-d\ell} < |\v{k}| < \Lambda$ near the UV momentum cutoff $\Lambda$. This is followed by the conventional rescaling of momenta, frequency and fields. To restore the original cutoff 
we rescale momenta as $\v{k}=\v{k}' e^{-d\ell}$ while $\omega=\omega' e^{-z d\ell}$ with $z$ the dynamical exponent.  The fields are rescaled as
\begin{equation}
\begin{gathered}
\Psi(\v{k},\omega)  = \Psi'(\v{k}',\omega') e^{-\delta_\Psi d\ell/2},\\
\phi(\v{k},\omega)  = \phi'(\v{k}',\omega') e^{-\delta_\phi d\ell/2}.
\end{gathered}
\end{equation}

We start with a simple tree-level scaling analysis. In the absence of disorder and at the critical point $m^2=0$ the field theory remains invariant under the above rescaling for $z=1$, $\delta_\Psi = -2-2z$, 
and $\delta_\phi=-4$. As the tuning parameter of the quantum phase transition, the order-parameter mass is a relevant perturbation with tree-level scaling dimension $[m^2]=2-z$. On the other hand, the 
bosonic $\phi^4$ vertex (\ref{eq.phi4}) is irrelevant with scaling dimension $[\lambda] = -6 -3z-2\delta_\phi=-1$, justifying why we neglected it in our theory, Eq.~(\ref{eqn:full_replica_action}). Under these 
scaling conventions the fermionic disorder is vertex is marginal at tree level which motivates a perturbative expansion in the couplings $\Delta_i$ of fermionic disorder.

\begin{figure}[t]
\includegraphics[width=0.7\columnwidth]{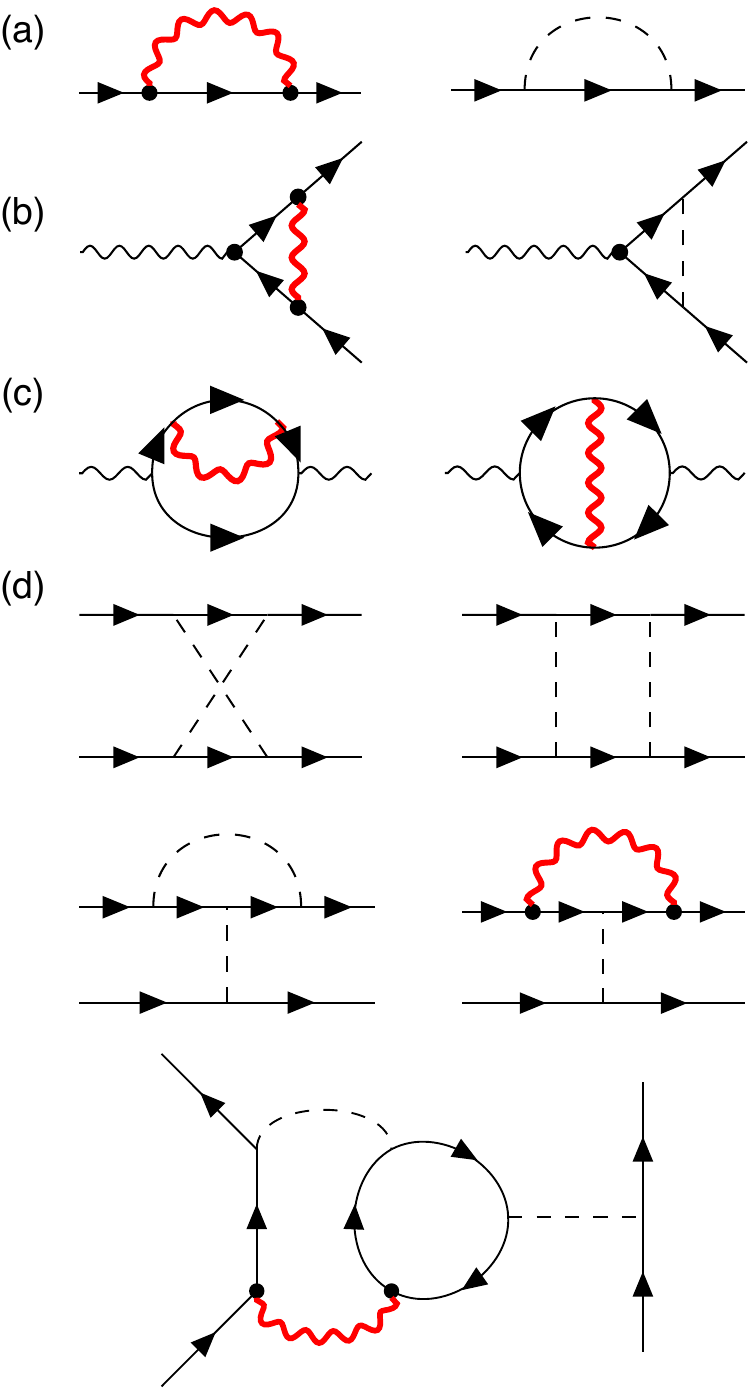}
\caption{Feynman diagrams of $\mathcal{O}(\Delta_i,\frac{1}{N})$ for the large-$N$ theory, Eq.~(\ref{eqn:full_replica_action}). (a) Fermion self energy corrections that renormalize the fermionic propagator. (b) Renormalization of the Yukawa vertex. (c) Renormalization of the bosonic mass.  (d) Corrections to the fermionic disorder vertex. 
 The dashed line represents the replicated disorder interaction. }
\label{fig:diagrams}
\end{figure}

We compute all diagrams, shown in Fig.~\ref{fig:diagrams}, that contribute at $\mathcal{O}(\Delta_i,\frac{1}{N})$ in $d=2$. We first consider the fermionic self-energy corrections due to the Yukawa 
coupling at second order and the disorder vertex at linear order, which are shown by the two diagrams in Fig.~\ref{fig:diagrams}(a), respectively. The first diagram leads to a renormalization of
the overall prefactor of the inverse fermion propagator, resulting in an anomalous dimension of the fermion fields.  The disorder induced self energy only affects the frequency term and therefore breaks
the symmetry between momentum and frequency scaling, leading to a correction to the dynamical exponent $z$. The inverse fermion propagator remains invariant under the RG for
\begin{align}
\delta_\Psi &= -4+  \frac{8}{3 \pi^2 N } - \frac12 \left(\tilde{\Delta}_0+\tilde{\Delta}_z+2\tilde{\Delta}_\perp\right),
\label{eqn:fermion_critical_dimension}\\
z & = 1 + \frac12\left(\tilde{\Delta}_0+\tilde{\Delta}_z+2\tilde{\Delta}_\perp\right),
\label{eqn:dynamical_exponent}
\end{align}
where we have defined the rescaled disorder variances 
\begin{equation}
\tilde{\Delta}_i = \frac{\Delta_i}{\pi v^2} 
\end{equation}
and $\tilde{\Delta}_\perp := \tilde{\Delta}_x = \tilde{\Delta}_y$. Hence the theory will no longer be Lorentz invariant  for any finite disorder fixed point. The renormalization of the Yukawa 
vertex is calculated from the diagrams in Fig.~\ref{fig:diagrams}(b), 
\begin{align}
\frac{d g}{d\ell} &= \left[-4 -2z  -\delta_\Psi-\frac{\delta_\phi}{2}\right. \\ 
&\left. \quad - \frac{8}{\pi^2 N} +\frac14 \left(2\tilde{\Delta}_\perp-\tilde{\Delta}_z-\tilde{\Delta}_0\right) \right]g.\nonumber
\end{align}

Since the coupling $g$ can be scaled out of the large-$N$ replica theory, Eq.~(\ref{eqn:full_replica_action}), using $\phi \to \phi/g$, $m^2\to g^2 m^2$, we demand that it is scale invariant. This determines 
the critical dimension of the order-parameter field $\phi$,
\begin{equation}
\delta_\phi = -4 -\frac{64}{3 \pi^2 N} -\frac12 \left(2\tilde{\Delta}_\perp+3\tilde{\Delta}_z + 3\tilde{\Delta}_0\right), \label{eqn:bosonic_critical_dimension}
\end{equation} 
where we have eliminated $\delta_\Psi$ and $z$, using Eqs. (\ref{eqn:fermion_critical_dimension}) and (\ref{eqn:dynamical_exponent}). 
The renormalization of the order parameter mass $m^2$ is given by the two-loop diagrams in Fig.~\ref{fig:diagrams}(c). The resulting RG equation is given by
\begin{equation}
\frac{d m^2}{d\ell} =\left( 1 -\frac{32}{3 \pi^2 N} + \tilde{\Delta}_z + \tilde{\Delta}_0\right) m^2, \label{eqn:rg_flow_bosonic_mass}
\end{equation}
where the dependence on disorder arrises through $z$ (\ref{eqn:dynamical_exponent}) and $\delta_\phi$ (\ref{eqn:bosonic_critical_dimension}). At any finite disorder 
fixed point the order parameter correlation length exponent $\nu$, defined through the identification $d m^2/d\ell = \nu^{-1}m^2$, will therefore differ from the one in the 
clean system.

Finally, the coupled RG equations for the different types of disorder are obtained from the diagrams in Fig.~\ref{fig:diagrams}(d), 
\begin{align}
\frac{d\tilde{\Delta}_0}{d\ell} &= \tilde{\Delta}_0 \left(\tilde{\Delta}_0+\tilde{\Delta}_z+2\tilde{\Delta}_\perp - \frac{32}{9 \pi^2 N}\right) + 2 \tilde{\Delta}_\perp \tilde{\Delta}_z,\nonumber\\
\frac{d\tilde{\Delta}_\perp}{d\ell} &=  -\tilde{\Delta}_\perp\left(\frac{\tilde{\Delta}_z}{6} +   \frac{32}{9\pi^2 N}  \right)+\tilde{\Delta}_0\tilde{\Delta}_z,  
\label{eqn:rg_flow_disorder}\\
\frac{d\tilde{\Delta}_z}{d\ell} &= \tilde{\Delta}_z\left(\frac{5\tilde{\Delta}_\perp}{3} - \tilde{\Delta}_z-\tilde{\Delta}_0-\frac{32}{3 \pi^2 N}\right)+ 2\tilde{\Delta}_\perp\tilde{\Delta}_0.\nonumber
\end{align}

In the non-interacting limit, corresponding to diagrams in Fig.~\ref{fig:diagrams} that only include the disorder vertex,  the RG equations for the disorder variances agree with previous
 results \cite{grinstein_qhe_disorder, ye_weak_disorder_quantum_hall,mirlin_transport_disordered_graphene,aleiner_graphene_conductivity}.

\subsection{RG flow and fixed points}

We start by summarizing the critical exponents for the interaction-driven semimetal to CDW insulator transition of the clean system at $T=0$ in $d=2$. The critical exponents of order $1/N$ at the clean interacting critical fixed point, which we denote by $P_{\text{clean}}$, are obtained from Eqs. (\ref{eqn:fermion_critical_dimension}), (\ref{eqn:dynamical_exponent}), (\ref{eqn:bosonic_critical_dimension}) and 
(\ref{eqn:rg_flow_bosonic_mass}) by setting $\tilde{\Delta}_0 = \tilde{\Delta}_\perp = \tilde{\Delta}_z =0$.
In the absence of disorder the theory satisfies Lorentz invariance with dynamical exponent $z=1$. The anomalous critical dimensions of the 
fields, defined through $\delta_\Psi = -4 +\eta_\Psi$, $\delta_\phi = -4 +\eta_\phi$, and the correlation length exponent $\nu$ reduce to
\begin{equation}
\begin{gathered}
\eta_\Psi^\text{clean} = \frac{8}{3\pi^2 N},\quad \eta_\phi^\text{clean} = -\frac{64}{3\pi^2 N},\\
\nu_\text{clean} = 1 +\frac{32}{3\pi^2 N}.
\end{gathered}
\end{equation}

These exponents are in agreement with those obtained from soft cutoff RG \cite{uryszek_soft_cut_off} and with previous results using the large $N$ conformal bootstrap 
\cite{Vasilev_nu_n2_1993,Gracey94,Iliesiu+18} and the critical point large $N$ formalism \cite{Gat1990,Gracey_eta_n2_1991,Gracey_nu_n2_1992}. At $P_\text{clean}$, the order parameter 
mass $m^2$ is the only relevant parameter, representing the tuning parameter of the quantum phase transition. 

\begin{figure}[t]
\includegraphics[width=\columnwidth]{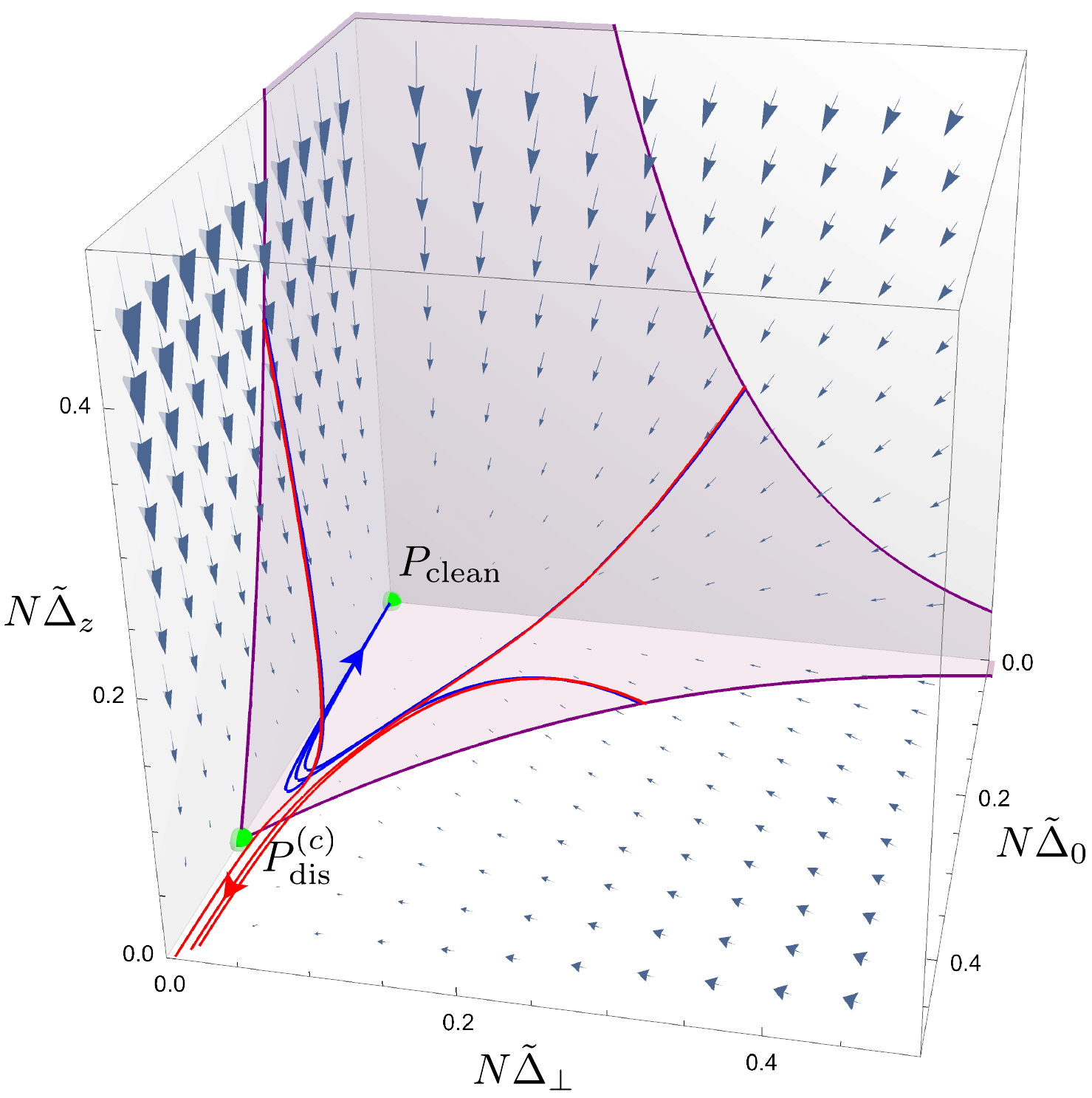}
\caption{RG flow in the disorder subspace on the critical manifold $m^2=0$, as defined by Eqs. (\ref{eqn:rg_flow_disorder}). Within the region bounded by the transparent surface disorder 
renormalizes to zero, showing 
that the CDW critical point $P_\text{clean}$ is stable against small disorder. Near this boundary surface the RG flow is towards a finite disorder fixed point $P_\text{dis}^{(c)}$ at which only chemical 
potential disorder is relevant.}
\label{fig:rg_flow}
\end{figure}

In order to analyze whether the clean system CDW critical point is stable against weak charge-impurity disorder, we numerically integrate the coupled 
RG equations for $\tilde{\Delta}_0$, $\tilde{\Delta}_\perp$ and $\tilde{\Delta}_z$ (\ref{eqn:rg_flow_disorder}). The resulting RG flow of the three disorder variances on the critical manifold $m^2=0$ is 
shown in Fig.~\ref{fig:rg_flow}. For small disorder, in the regime bounded by the transparent purple surface, the flow is towards the clean system critical point $P_\text{clean}$, demonstrating 
that the CDW quantum critical point is stable against weak disorder.  This is in line with the Harris criterion which states that 
a non-disordered fixed point is stable if
$\nu_\text{clean}\ge 2/d$, where $d$ is the dimensionality of the system \cite{Harris74,Chayes+86}. 

Close to the boundary surface, the RG flow is controlled by the only finite disorder fixed point in the accessible region of positive variances, 
\begin{equation}
P_\text{dis}^{(c)}:\;\left(\tilde{\Delta}_0^{(c)},\tilde{\Delta}_{\perp}^{(c)},\tilde{\Delta}_z^{(c)}\right) = \left(\frac{32}{9\pi^2 N},0,0\right).
\end{equation}

 $P_\text{dis}^{(c)}$ is unstable along the $\tilde{\Delta}_0$ direction but stable against $\tilde{\Delta}_\perp$ and $\tilde{\Delta}_z$. This is consistent with the RG flow for initial values $\tilde{\Delta}_i(0)$
 that are very close to the separating surface in Fig.~\ref{fig:rg_flow}. Shown are three pairs of trajectories with initial values that are slightly inside (blue) and outside (red) of the region bounded by the 
 surface. The trajectories closely track the surface and split very close to  $P_\text{dis}^{(c)}$, where the flow is either to the clean fixed point, $\tilde{\Delta}_0\to 0$ or strong chemical potential 
 disorder, $\tilde{\Delta}_0\to \infty$.

Our RG analysis shows that the transition to a glassy state is always driven by potential disorder, even if the other forms of disorder initially dominate. Since 
the random gauge field and random mass disorders are irrelevant at $P_\text{dis}^{(c)}$ we neglect them in the following. The RG equations for the  chemical potential disorder $\tilde{\Delta}_0$ and the order 
parameter mass $m^2$ then reduce to  
\begin{align}
\frac{d\tilde{\Delta}_0}{d\ell} &= \tilde{\Delta}_0 \left(\tilde{\Delta}_0- \frac{32}{9 \pi^2 N}\right),\\
\frac{d m^2}{d\ell} &=\left( 1 -\frac{32}{3 \pi^2 N} + \tilde{\Delta}_0\right) m^2.
\label{eqn:rg_flow_bosonic_mass2}
\end{align}  

Inserting the critical disorder strength $\tilde{\Delta}_0^{(c)} = \frac{32}{9 \pi^2 N}$ into Eqs. (\ref{eqn:fermion_critical_dimension}), (\ref{eqn:dynamical_exponent}), (\ref{eqn:bosonic_critical_dimension})
and (\ref{eqn:rg_flow_bosonic_mass2}) we obtain the critical exponents at the finite disorder multi-critical point $P_\text{dis}^{(c)}$,
\begin{equation}
\begin{gathered}
\eta_\Psi^\text{dirty} = \frac{8}{9\pi^2 N},\quad \eta_\phi^\text{dirty} = -\frac{80}{3\pi^2 N},\\
\nu_\text{dirty} = 1 +\frac{64}{9\pi^2 N},\quad z_\text{dirty} =1 + \frac{16}{9\pi^2 N}.
\end{gathered}
\end{equation}

At both the clean system semimetal to CDW insulator transition and at the finite disorder multicritical point the fermion anomalous dimension 
$\eta_\Psi$ is greater than zero. This implies that at the quantum critical points (QCPs) the fermion Green's function has branch cuts rather than quasiparticle poles,
and the fermionic liquid is therefore a non-Fermi liquid. Approaching the QCPs from the metallic side, $V<V_c$ and $V_c-V\to 0$, the quasiparticle residue has 
to vanish with some characteristic exponent. On the CDW side, the condensation of the order parameter leads to the formation of a gap $M$ in the fermion spectrum, 
which increases as a power of $V-V_c>0$. 

In order to extract these exponents we perform a scaling analysis of  the fermionic spectral function. Details can be found in Ref.~\cite{Herbut+09}. Here we only give the results.
Approaching the quantum phase transition from the metallic side, the quasiparticle pole strength vanishes as 
\begin{equation}
Z \sim (V_c - V)^{(z-1+\eta_\Psi)\nu} = (V_c - V)^{\frac{8}{3\pi^2N}},
\end{equation}
where to order $1/N$ the critical exponents are the same for the clean and dirty fixed points $P_\text{clean}$ and $P_\text{dirty}$. The Fermi velocity behaves as 
\begin{equation}
v \sim |V_c - V|^{(z-1)\nu} = \left\{ \begin{array}{cc} \text{const} & \text{at}\;P_\text{clean} \\ |V_c - V|^{\frac{16}{9\pi^2N}} & \text{at}\;P_\text{dirty} \end{array}\right.
\end{equation}

Finally, on the CDW insulator side of the quantum phase transition the gap in the electron spectrum increases as
\begin{equation}
M \sim (V - V_c)^{z\nu} = \left\{ \begin{array}{cc} (V-V_c)^{1+\frac{32}{3\pi^2N}} & \text{at}\;P_\text{clean} \\ (V- V_c)^{1+\frac{80}{9\pi^2N}} & \text{at}\;P_\text{dirty} \end{array}\right. 
\label{eq.gapM}
\end{equation}

\begin{figure}[t]
\includegraphics[width=0.95\columnwidth]{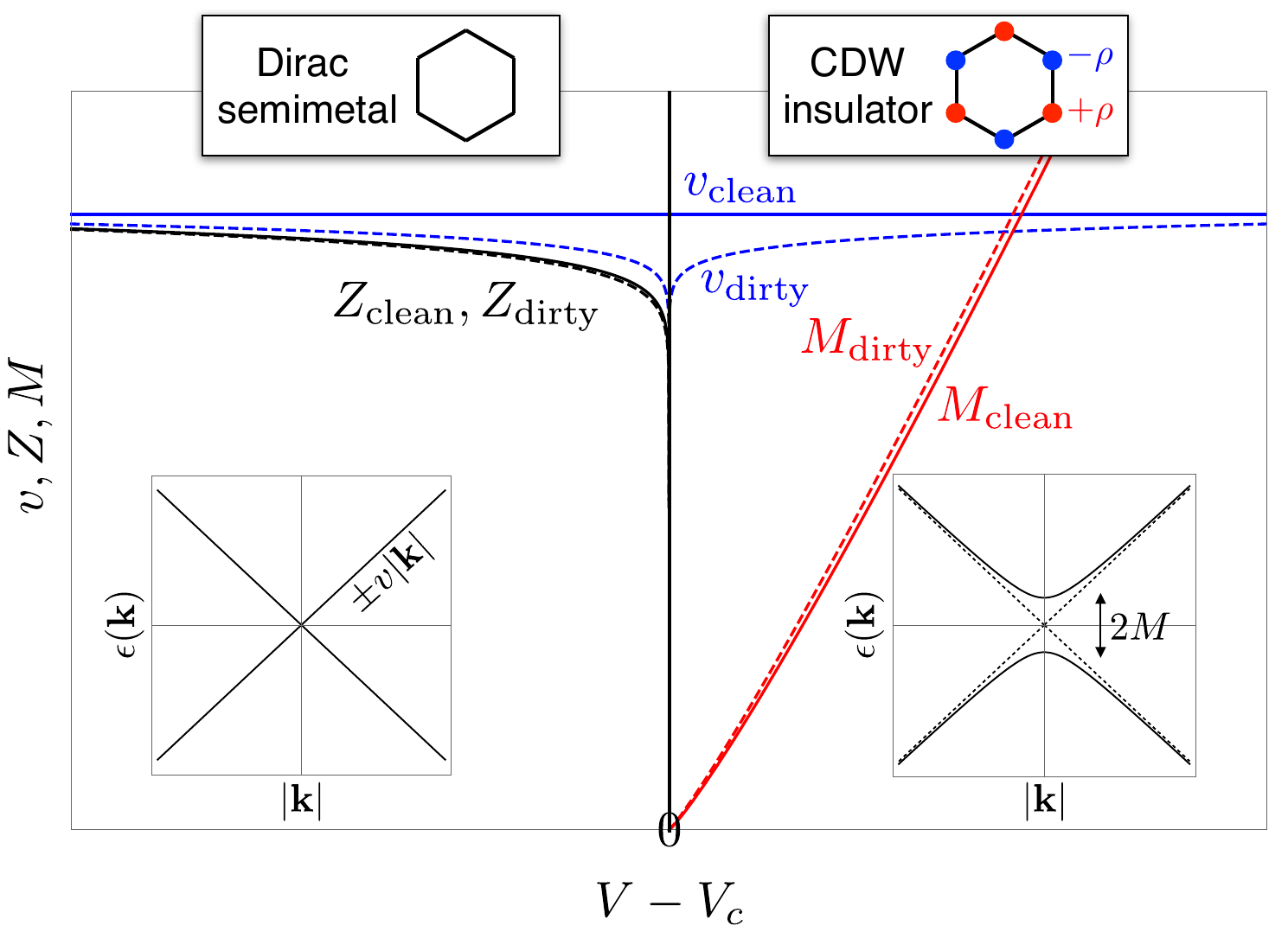}
\caption{Behavior of the the quasiparticle pole strength $Z$, the Fermi velocity $v$ and the gap $M$ in the fermion spectrum at the clean semimetal/CDW insulator transition and 
at the finite disorder multicritical point, as a function of the nearest neighbor repulsion $V-V_c$. Here we evaluated the critical exponents for $N=8$, corresponding to Dirac electrons on the 
honeycomb lattice with valley and spin degeneracies.}
\label{fig:MI_trans}
\end{figure}

The behavior of $Z$, $v$ and $M$ near the clean and finite-disorder QCPs is illustrated in Fig.~\ref{fig:MI_trans}.

In order to estimate the phase boundary between the CDW insulator and the disordered phase  in the close proximity of $P_\text{dirty}$ for $V>V_c$ 
and $\tilde{\Delta}_0>\tilde{\Delta}_0^{(c)}$ we compare the CDW induced gap $M$ in the electron spectrum with the standard deviation $\sqrt{\tilde{\Delta}_0}$ of the chemical 
potential disorder. Close to $P_\text{dirty}$, the disorder increase exponentially under the RG, 
\begin{equation}
\tilde{\Delta}_0 (\ell) - \tilde{\Delta}_0^{(c)} \simeq \left( \tilde{\Delta}_0 - \tilde{\Delta}_0^{(c)}\right) e^{\nu_\Delta^{-1}\ell}\; \text{with} \;\; \nu_\Delta^{-1}=\frac{32}{9\pi^2N}.\nonumber
\end{equation}

We evaluate the disorder variance at the ``correlation length" $\xi\sim e^{\ell^*}\sim (V-V_c)^{-\nu}$, where $m^2(\ell^*)\simeq -1$. Equating the resulting standard deviation with the gap $M$ 
near $P_\text{dirty}$, Eq.~(\ref{eq.gapM}), we obtain the phase boundary

\begin{align}
\left( \tilde{\Delta}_0 - \tilde{\Delta}_0^{(c)}\right) &\simeq (V-V_c)^{(2z_\text{dirty}+\nu_\Delta^{-1})\nu_\text{dirty}} \nonumber \\
&\simeq (V-V_c)^{2\left(1+\frac{32}{3\pi^2 N}  \right)}.
\end{align}

\begin{figure}[t]
\includegraphics[width=\columnwidth]{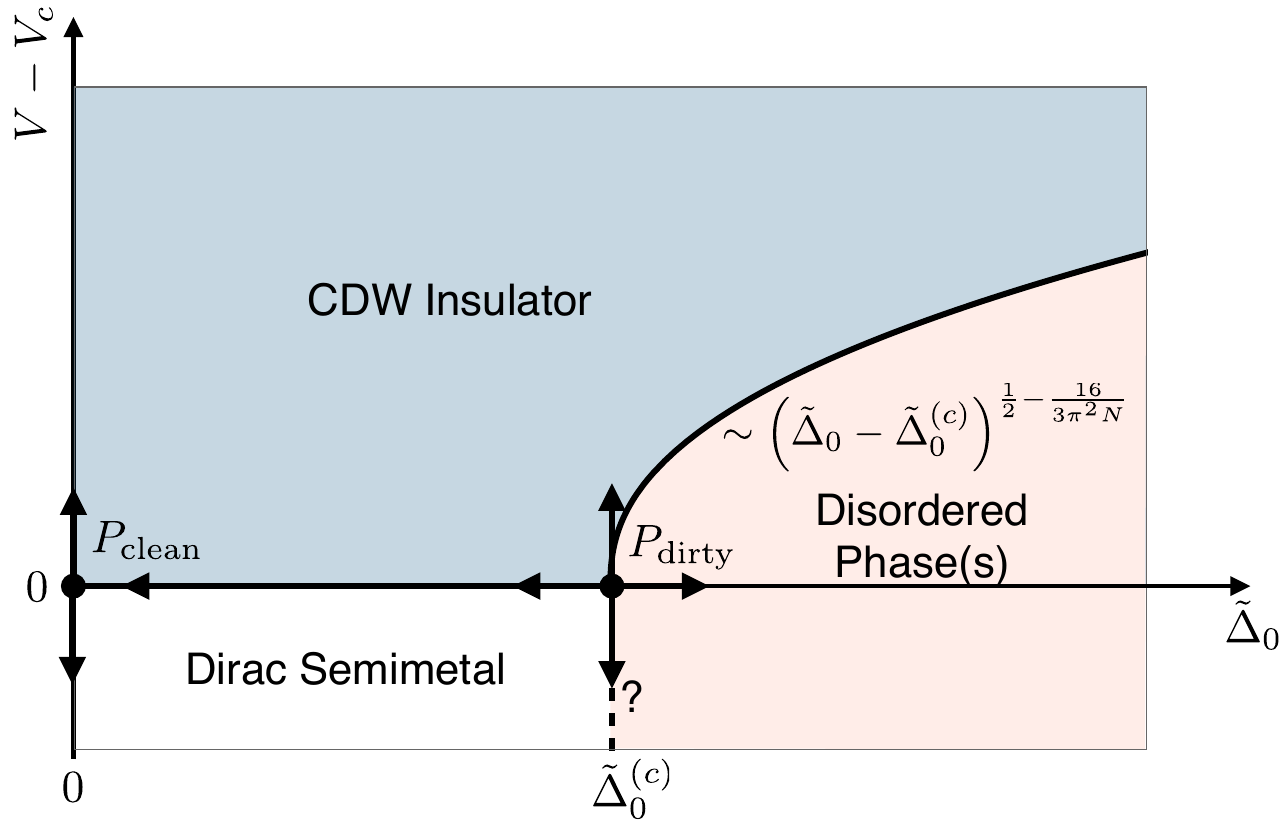}
\caption{Schematic Phase diagram as a function of the interaction strength $V-V_c$ and the variance $\tilde{\Delta}_0$ of the chemical potential disorder.}
\label{fig.phase}
\end{figure}

A schematic phase diagram as a function of the interaction strength $V-V_c \simeq -m^2$  and the chemical potential disorder $\tilde{\Delta}_0$ is shown in Fig.~\ref{fig.phase}.

\section{Discussion}\label{sec:discussion}

We have investigated the effects of quenched short-ranged disorder on the quantum phase transition between a two-dimensional Dirac semi-metal and a charge density wave (CDW) insulator. 
In the absence of disorder, the phase transition belongs to the chiral Ising Gross-Neveu-Yukawa (GNY) universality class. In order to achieve analytic control in $d=2$, far below the upper critical dimension, 
we have analyzed the problem in the limit of a large number $N$ of Dirac fermion flavors. 
We have used the RPA fermion loop resummation to self-consistently account for the Landau damping of the boson dynamics by electronic particle-hole excitations. As pointed out in the literature 
\cite{uryszek_soft_cut_off,chubukov_landau_damping},  this is a non-perturbative effect in two spatial dimensions that changes the IR physics and hence the universal critical behavior. As we 
have demonstrated in our work, Landau damping also plays a crucial  role in how the critical system responds to disorder. 

We have considered three types of electronic disorder that all arise from non-magnetic charge impurities. The random potential from the impurities is decomposed into  
random mass disorder, which locally breaks the symmetry between the two sublattices, and symmetric random chemical potential disorder. The local lattice deformations caused by impurity atoms 
is accounted for by random gauge potential disorder  \cite{mirlin_transport_disordered_graphene}. For simplicity, we have neglected correlations between the different types of disorder and assumed
that disorder is uncorrelated between different positions in space.  

After averaging over disorder, using the replica formalism, we have performed a perturbative RG calculation to leading order in the disorder strength and in $1/N$. Our analysis shows that 
the clean GNY critical point is stable against weak disorder. This is in stark contrast to non-interacting or weakly interacting two-dimensional Dirac fermions where disorder 
is a relevant perturbation, resulting in a run-away flow towards strong disorder \cite{efetov_transport_disorder_graphene,sachdev_coulomb_quantum_hall,ye_weak_disorder_quantum_hall,vozmediano_disorder_graphene}.
 
Most importantly, we have identified a dirty GNY critical point at a finite value of the chemical potential disorder of order $1/N$. At this multicritcal point, the random mass and random 
gauge potential disorders are irrelevant. This shows that the transition into a disordered state is driven by chemical potential disorder, even if the other forms of disorder dominate on shorter 
length and time scales.

The irrelevance of random mass disorder $V_z$  at the clean and finite disorder GNY critical points might seem surprising since this type of disorder breaks the AB sub-lattice symmetry, similar to a random field that 
couples to the Ising CDW order parameter $\phi$. According to the scaling arguments by Imry/Ma \cite{Imry+75,Aharony+76} and Aizenman/Wehr \cite{Aizenman+89}, such a random field would destroy any long-range Ising order and associated quantum 
critical point in two spatial dimensions. However, the GNY universality class falls outside a pure order-parameter description used in these arguments,  and is formulated in terms of bosonic and fermionic degrees 
of freedom. The coupling to gapless fermion excitations changes the IR dynamics of the bosonic order parameter field, resulting in unusual scaling properties in the bosonic sector, e.g. the $\phi^4$ vertex is rendered 
irrelevant in $d=2$. Moreover, the disorder $V_z$ couples to the fermion operator $\Psi^\dagger \sigma^z \Psi$ and not to the CDW order parameter $\phi$. Although after integrating out the fermions, $V_z$ would 
translate into a random field in the resulting order parameter theory, this step is not allowed  since the fermions are gapless at the QCP. Future quantum Monte-Carlo studies could shed some light on this subtle question.

 The disorder driven phase transition along the line of critical interaction in the two-dimensional system might be similar to the transition in weakly interacting, three dimensional 
Weyl/Dirac semimetals \cite{chakravarty_diffusive_metal,gonzalez_3d_weyl_coulomb_disorder,das_sarma_disorder_3d_dirac,wang_weyl_3d_coulomb_disorder,Syzranov+18}. In both cases, the transition is 
driven by chemical potential disorder which is expected to induce a finite zero-energy density of states in the disordered phase, giving rise to diffusive metallic behavior. 
This would be consistent with our naive picture for the transition between the CDW insulator, which forms above the critical interaction strength, and the disordered phase: if the standard deviation of
the random chemical potential shifts exceeds the electronic gap induced by the symmetry breaking, the system will develop a finite density of states at the average chemical potential, leading to 
diffusive metallic behavior. However, further calculations are required to ascertain the properties of the disordered phase in the strongly interacting, two dimensional system. An investigation of  
the dependence on the form of the disorder distribution, e.g. whether it is bounded, Gaussian or exhibits long tails, as well as of any potential replica symmetry 
breaking \cite{parisi_replica_order_parameter_1979}, indicative of glassy behavior, would be very interesting. The random-mass disorder might play an important role in stabilizing a
finite disorder multi-critical point with broken Replica symmetry.
 
Our renormalization-group approach  does not capture non-perturbative, rare region effects, which have spurned a lot of discussion in the context of three dimensional Weyl/Dirac semimetals.
A study by Nandkishore \textit{et al.} \cite{sondhi_rare_region_effects_dirac_3D} first proposed that rare region effects induce a non-vanishing density of states at the Weyl/Dirac points, thereby 
turning the disorder-driven phase transition into a crossover. This was substantiated by numerical calculations \cite{pixley_avoided_QCP_weyl,Pixley+16} but remains at odds with recent theoretical 
literature \cite{altland_dos_weyl_rare_region,altland_weyl_stable_against_disorder}. However, as chemical potential disorder is marginal in two spatial dimensions, and irrelevant in three, 
it is expected that rare region resonances will have a ``sub-leading effect'' on the physics of the transition in two dimensions \cite{pixley_rare_region_effects_review}.

We have shown that the symmetry-breaking quantum phase transition at the dirty GNY does not belong to the chiral-Ising GNY universality of the clean system. We have computed 
 the critical exponents at the finite-disorder multi-critical point to order $1/N$ and found that the anomalous dimensions of the boson and fermion fields, the correlation length 
exponent of the CDW order parameter and the dynamical critical exponent differ from those at the clean GNY fixed point. This leads to different critical behavior of physical observables 
such as the electronic gap, the Fermi velocity, and the quasi-particle residue near the transition and results in a novel non-Fermi liquid state at the multicritical point.

The interplay between symmetry breaking and disorder was previously studied for the XY GNY \cite{sondhi_sc_of_disordered_Dirac,maciejko_dirac_disorder_double_epsilon} and the chiral 
Ising and Heisenberg GNY models \cite{maciejko_gny_random_mass_disorder}, using the replica 
formalism combined with $\epsilon$ expansions. Near the upper critical dimension fermionic disorder is strongly irrelevant at the clean system quantum critical points. 
Instead, short-ranged disorder of the bosonic order parameter mass (sometimes referred to as random $T_{\text{c}}$ disorder)  gives rise to a finite disorder multicritical point, regardless of the 
symmetry of the order parameter. At this finite disorder critical point the Lorentz invariance is broken with a dynamical exponent $z>1$, similar to our dirty GNY fixed point, while the fermionic and 
bosonic anomalous dimensions remain unchanged, which is not the case in our theory.   
  
The irrelevance of the chemical potential disorder seems to be only valid near the upper critical dimension, hence any extrapolation to the physical dimension of $d=2$ without the inclusion of it is 
 questionable. Moreover, the non-perturbative Landau damping which is crucial for the universal critical behavior of the two-dimensional system, is not captured by an $\epsilon$ expansion below the 
 upper critical dimension.   On the other hand, we have not included bosonic disorder in our effective field theory, for simplicity.  Starting from an interacting fermionic model with a 
 random potential, bosonic disorder would not arise from a Hubbard-Stratonovich decoupling of the fermionic interaction vertex.  However, as pointed out in Refs. \cite{sondhi_sc_of_disordered_Dirac} and \cite{maciejko_dirac_disorder_double_epsilon}, at two-loop order chemical potential disorder could generate a bosonic disorder 
 vertex in the replica theory. We have presented an explicit calculation 
 in Appendix \ref{appx:bosonic_disorder_generation}, demonstrating that this is not the case.

It is important to stress, however, that bosonic random mass and random field disorders do not break symmetries that are not already broken by the fermionic disorder potentials. An effective low energy 
field theory obtained from careful coarse-graining of a microscopic lattice Hamiltonian will therefore also contain the symmetry allowed bosonic disorder. The presence of additional bosonic disorder could 
potentially affect our conclusions and should be considered in future work. Random field disorder is known to have a detrimental effect on the CDW order and associated quantum phase transition. 
However, in the case of remote charge impurities that do not break the symmetry between the two sublattices, such random field disorder would  be suppressed. As the fermionic random potentials, the bosonic 
random mass disorder is marginal at the GNY interacting fixed point and might therefore alter the multi-critical behavior.

It is also interesting to compare our results with recent work \cite{Zhao+17,Goswami+17,Thomson+17} on the role of generic types of fermionic disorder in strongly coupled QED$_3$, which describes 
the interaction of massless Dirac fermions  with U(1) gauge bosons in 2+1 space-time dimensions. Similar to our work, the problem was generalized to a large number $N$ of fermion flavors and  
analyzed within the Replica framework. In QED$_3$, sufficiently strong gauge coupling leads to dynamical chiral symmetry breaking and spontaneous fermion mass generation. However, this quantum 
phase transition is lost above a critical number $N_c=32/\pi^2$ of fermion flavors \cite{Appelquist+86,Appelquist+88,Nash89,Maris96} and therefore no longer accessible in the large $N$ 
limit \cite{Zhao+17}. This might explain why disorder is found to be a relevant perturbation, similar to the case of weakly interacting Dirac fermions in 2+1 dimensions \cite{sachdev_coulomb_quantum_hall,ye_weak_disorder_quantum_hall,vozmediano_disorder_graphene}. However, unlike in the weakly interacting case, the flow is towards a stable finite-disorder 
fixed point with a broken flavour degeneracy and $z>1$. This behaviour is very different from that of  chiral Ising GNY theories at criticality, investigated in our work: the clean GNY fixed point in 2+1 
dimensions is stable against weak fermionic disorder and the transition to a diffusive metallic state is characterized by a multi-critical point at finite chemical potential disorder and $z>1$. 

For simplicity, we have analyzed critical GNY theories with an Ising order parameter. We believe that the behaviour is similar for GNY theories with continuous order parameter symmetries 
and that the stability of the clean GNY fixed point against disorder is the consequence of gapless fermion excitations that completely change the long-wavelength order-parameter dynamics 
in two spatial dimensions. The Wilson-Fisher critical fixed point in conventional bosonic theories, e.g. for the superfluid-insulator transition $d=2$, is indeed unstable towards the formation of a 
finite disorder fixed point \cite{Goldman+20}. Although the behaviour of the large $N$ field theory in $d=2$ is similar to that of a double $\epsilon$ expansion near the upper critical dimension, the latter shows 
a spiralling RG flow into the finite disorder fixed point \cite{Goldman+20}.  Similar behaviour is found in a double $\epsilon$ expansion of critical GNY theories with bosonic random mass 
disorder \cite{maciejko_dirac_disorder_double_epsilon}. This could either point towards pathologies of the double $\epsilon$ expansion or otherwise indicate important physical behaviour that is 
lost in the oversimplified 
large $N$ treatment.

In future extensions of our work it would be interesting to investigate the effects of long-range correlations of disorder. It is often assumed that impurities and imperfections are screened effectively 
and that disorder can therefore be taken to be uncorrelated. However,  it has been reported that in graphene the correlations between disorder-induced puddles of electron- and hole-doped regions 
decay algebraically \cite{martin_puddles_graphene_stm,deshpande_spectroscopyt,zhang_graphene_inhomogeneity}. Such power-law correlations are expected to change the long-wavelength physics 
and hence the universal critical behavior. One might also include other types of disorder, e.g. defects that lead to inter-valley scattering, magnetic impurities that break the spin degeneracy, or topological 
lattice defects that are described by random non-Abelian gauge fields. The interplay of the different types of disorder is expected to lead to rich phase behavior and novel critical phenomena, in particular if 
competing fermionic interactions are taken into account.

\section*{Acknowledgments}
F.K. acknowledges financial support from EPSRC under Grant No. EP/P013449/1.

\appendix
\section{Two-loop fermion diagram that generates the boson disorder}\label{appx:bosonic_disorder_generation}

\begin{figure}[t!]
\includegraphics[width=0.55\columnwidth]{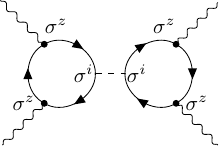}
\caption{The two-loop diagram that according to Refs.\cite{sondhi_sc_of_disordered_Dirac,maciejko_dirac_disorder_double_epsilon} generates the bosonic disorder vertex.}
\label{fig:boson-two-loop-disorder}
\end{figure}

Here we address the question if the electronic disorder, which are defined on the level of the quadratic fermion action [see Eq.~(\ref{eq.dis})], can generate 
random mass disorder of the bosonic order parameter field  at two loop order, as suggested in Refs. \cite{sondhi_sc_of_disordered_Dirac} and \cite{maciejko_dirac_disorder_double_epsilon}.
In the disorder averaged replica theory the electronic disorder is described by a disorder vertex that is quartic in the fermionic Grassmann fields, couples different replicas, and is non-local in imaginary time
[see Eq.~(\ref{eqn:full_replica_action})]. 
Similarly, bosonic random mass disorder gives rise to a disorder vertex
\begin{equation}
S_\phi^\textrm{dis} = - \frac{\sigma^2}{2}\sum_{a,b=1}^n \int d^2\v{x} \int d\tau \int d\tau^\prime \phi^2_{a}(\v{x},\tau) \phi^2_{b}(\v{x},\tau')
\end{equation}
in the replica field theory, where $\sigma^2$ is the variance of the bosonic random mass disorder distribution. This vertex would be generated by the two-loop diagram shown in Fig.~\ref{fig:boson-two-loop-disorder} where the external momenta in the loop integrals are set to zero. This results in 
\begin{equation}
\sigma^2 \sim \frac{g^4}{N^2}\sum_{i=0,x,y,z} D_i^2 \Delta_i
\end{equation}
with 
\begin{equation}
D_i =  \int_{\v{k},\omega}\Tr\left[G_\Psi(\v{k},\omega)\sigma^z G_\Psi(\v{k},\omega) \sigma^z G_\Psi(\v{k},\omega) \sigma^i \right].
\end{equation} 

It is straightforward to see that electronic random mass disorder $\Delta_z$ does not contribute since the trace over the product of Pauli matrices vanishes in this case, $D_z=0$. In the other 
channels we obtain the integrals
\begin{align*}
D_0 &= -N\int_{\v{k},\omega} \ \frac{ i \omega}{(\omega^2 + v^2\v{k}^2)^2}, \\
D_x &= -N\int_{\v{k},\omega}    \ \frac{v k_x}{(\omega^2 + v^2\v{k}^2)^2}, \\
D_y &= -N\int_{\v{k},\omega}  \ \frac{v k_y}{(\omega^2 + v^2\v{k}^2)^2},
\end{align*}
after taking the trace.  These integrals are either odd in the frequency or momenta and therefore evaluate to zero. This shows that 
for the chiral Ising GNY theory with purely electronic disorder, the bosonic disorder vertex is not generated at two-loop order.

There are certain higher-loop diagrams that vanish for similar reasons, or after taking the Replica limit $n\to 0$. However, we don't see a general argument for why boson mass disorder 
can't be generated at higher-loop order. As stated in the main text, there are other ways to generate boson mass disorder, e.g. by considering disorder in the nearest neighbour fermion interaction 
 before Hubbard-Stratonovich decoupling.
 
 We stress that the two-loop diagrams only vanish if external frequencies and momenta are set to zero. But only such diagrams result in a boson vertex of the form $\phi_a^2 \phi_b^2$ corresponding 
to Replica averaged random mass disorder. Expanding out external momenta is equivalent to a gradient expansion and gives rise to additional boson vertices such  $(\nabla \phi_a)^2 \phi_b^2$ or 
$(\nabla \phi_a)^2 (\nabla \phi_b)^2$, which are 
irrelevant under the RG.


\begin{thebibliography}{74}%
\makeatletter
\providecommand \@ifxundefined [1]{%
 \@ifx{#1\undefined}
}%
\providecommand \@ifnum [1]{%
 \ifnum #1\expandafter \@firstoftwo
 \else \expandafter \@secondoftwo
 \fi
}%
\providecommand \@ifx [1]{%
 \ifx #1\expandafter \@firstoftwo
 \else \expandafter \@secondoftwo
 \fi
}%
\providecommand \natexlab [1]{#1}%
\providecommand \enquote  [1]{``#1''}%
\providecommand \bibnamefont  [1]{#1}%
\providecommand \bibfnamefont [1]{#1}%
\providecommand \citenamefont [1]{#1}%
\providecommand \href@noop [0]{\@secondoftwo}%
\providecommand \href [0]{\begingroup \@sanitize@url \@href}%
\providecommand \@href[1]{\@@startlink{#1}\@@href}%
\providecommand \@@href[1]{\endgroup#1\@@endlink}%
\providecommand \@sanitize@url [0]{\catcode `\\12\catcode `\$12\catcode
  `\&12\catcode `\#12\catcode `\^12\catcode `\_12\catcode `\%12\relax}%
\providecommand \@@startlink[1]{}%
\providecommand \@@endlink[0]{}%
\providecommand \url  [0]{\begingroup\@sanitize@url \@url }%
\providecommand \@url [1]{\endgroup\@href {#1}{\urlprefix }}%
\providecommand \urlprefix  [0]{URL }%
\providecommand \Eprint [0]{\href }%
\providecommand \doibase [0]{https://doi.org/}%
\providecommand \selectlanguage [0]{\@gobble}%
\providecommand \bibinfo  [0]{\@secondoftwo}%
\providecommand \bibfield  [0]{\@secondoftwo}%
\providecommand \translation [1]{[#1]}%
\providecommand \BibitemOpen [0]{}%
\providecommand \bibitemStop [0]{}%
\providecommand \bibitemNoStop [0]{.\EOS\space}%
\providecommand \EOS [0]{\spacefactor3000\relax}%
\providecommand \BibitemShut  [1]{\csname bibitem#1\endcsname}%
\let\auto@bib@innerbib\@empty
\bibitem [{\citenamefont {Hasan}\ and\ \citenamefont
  {Kane}(2010)}]{hasan_kane_top_insulator_review}%
  \BibitemOpen
  \bibfield  {author} {\bibinfo {author} {\bibfnamefont {M.~Z.}\ \bibnamefont
  {Hasan}}\ and\ \bibinfo {author} {\bibfnamefont {C.~L.}\ \bibnamefont
  {Kane}},\ }\href {https://doi.org/10.1103/RevModPhys.82.3045} {\bibfield
  {journal} {\bibinfo  {journal} {Rev. Mod. Phys.}\ }\textbf {\bibinfo {volume}
  {82}},\ \bibinfo {pages} {3045} (\bibinfo {year} {2010})}\BibitemShut
  {NoStop}%
\bibitem [{\citenamefont {Vafek}\ and\ \citenamefont
  {Vishwanath}(2014)}]{vafek_vishwanath_diracF_in_materials_review}%
  \BibitemOpen
  \bibfield  {author} {\bibinfo {author} {\bibfnamefont {O.}~\bibnamefont
  {Vafek}}\ and\ \bibinfo {author} {\bibfnamefont {A.}~\bibnamefont
  {Vishwanath}},\ }\href
  {https://doi.org/10.1146/annurev-conmatphys-031113-133841} {\bibfield
  {journal} {\bibinfo  {journal} {Annual Review of Condensed Matter Physics}\
  }\textbf {\bibinfo {volume} {5}},\ \bibinfo {pages} {83} (\bibinfo {year}
  {2014})}\BibitemShut {NoStop}%
\bibitem [{\citenamefont {Armitage}\ \emph {et~al.}(2018)\citenamefont
  {Armitage}, \citenamefont {Mele},\ and\ \citenamefont
  {Vishwanath}}]{armitage_review_semimetals}%
  \BibitemOpen
  \bibfield  {author} {\bibinfo {author} {\bibfnamefont {N.~P.}\ \bibnamefont
  {Armitage}}, \bibinfo {author} {\bibfnamefont {E.~J.}\ \bibnamefont {Mele}},\
  and\ \bibinfo {author} {\bibfnamefont {A.}~\bibnamefont {Vishwanath}},\
  }\href {https://doi.org/10.1103/RevModPhys.90.015001} {\bibfield  {journal}
  {\bibinfo  {journal} {Rev. Mod. Phys.}\ }\textbf {\bibinfo {volume} {90}},\
  \bibinfo {pages} {015001} (\bibinfo {year} {2018})}\BibitemShut {NoStop}%
\bibitem [{\citenamefont
  {Fradkin}(1986{\natexlab{a}})}]{fradkin_semiconductor_disorder_I}%
  \BibitemOpen
  \bibfield  {author} {\bibinfo {author} {\bibfnamefont {E.}~\bibnamefont
  {Fradkin}},\ }\href {https://doi.org/10.1103/PhysRevB.33.3257} {\bibfield
  {journal} {\bibinfo  {journal} {Phys. Rev. B}\ }\textbf {\bibinfo {volume}
  {33}},\ \bibinfo {pages} {3257} (\bibinfo {year}
  {1986}{\natexlab{a}})}\BibitemShut {NoStop}%
\bibitem [{\citenamefont
  {Fradkin}(1986{\natexlab{b}})}]{fradkin_semiconductor_disorder}%
  \BibitemOpen
  \bibfield  {author} {\bibinfo {author} {\bibfnamefont {E.}~\bibnamefont
  {Fradkin}},\ }\href {https://doi.org/10.1103/PhysRevB.33.3263} {\bibfield
  {journal} {\bibinfo  {journal} {Phys. Rev. B}\ }\textbf {\bibinfo {volume}
  {33}},\ \bibinfo {pages} {3263} (\bibinfo {year}
  {1986}{\natexlab{b}})}\BibitemShut {NoStop}%
\bibitem [{\citenamefont {Aleiner}\ and\ \citenamefont
  {Efetov}(2006)}]{efetov_transport_disorder_graphene}%
  \BibitemOpen
  \bibfield  {author} {\bibinfo {author} {\bibfnamefont {I.~L.}\ \bibnamefont
  {Aleiner}}\ and\ \bibinfo {author} {\bibfnamefont {K.~B.}\ \bibnamefont
  {Efetov}},\ }\href {https://doi.org/10.1103/PhysRevLett.97.236801} {\bibfield
   {journal} {\bibinfo  {journal} {Phys. Rev. Lett.}\ }\textbf {\bibinfo
  {volume} {97}},\ \bibinfo {pages} {236801} (\bibinfo {year}
  {2006})}\BibitemShut {NoStop}%
\bibitem [{\citenamefont {Ostrovsky}\ \emph {et~al.}(2006)\citenamefont
  {Ostrovsky}, \citenamefont {Gornyi},\ and\ \citenamefont
  {Mirlin}}]{mirlin_transport_disordered_graphene}%
  \BibitemOpen
  \bibfield  {author} {\bibinfo {author} {\bibfnamefont {P.~M.}\ \bibnamefont
  {Ostrovsky}}, \bibinfo {author} {\bibfnamefont {I.~V.}\ \bibnamefont
  {Gornyi}},\ and\ \bibinfo {author} {\bibfnamefont {A.~D.}\ \bibnamefont
  {Mirlin}},\ }\href {https://doi.org/10.1103/PhysRevB.74.235443} {\bibfield
  {journal} {\bibinfo  {journal} {Phys. Rev. B}\ }\textbf {\bibinfo {volume}
  {74}},\ \bibinfo {pages} {235443} (\bibinfo {year} {2006})}\BibitemShut
  {NoStop}%
\bibitem [{\citenamefont {Ostrovsky}\ \emph {et~al.}(2007)\citenamefont
  {Ostrovsky}, \citenamefont {Gornyi},\ and\ \citenamefont
  {Mirlin}}]{mirlin_minimial_conductivity_quantum_critical_graphene}%
  \BibitemOpen
  \bibfield  {author} {\bibinfo {author} {\bibfnamefont {P.~M.}\ \bibnamefont
  {Ostrovsky}}, \bibinfo {author} {\bibfnamefont {I.~V.}\ \bibnamefont
  {Gornyi}},\ and\ \bibinfo {author} {\bibfnamefont {A.~D.}\ \bibnamefont
  {Mirlin}},\ }\href {https://doi.org/10.1103/PhysRevLett.98.256801} {\bibfield
   {journal} {\bibinfo  {journal} {Phys. Rev. Lett.}\ }\textbf {\bibinfo
  {volume} {98}},\ \bibinfo {pages} {256801} (\bibinfo {year}
  {2007})}\BibitemShut {NoStop}%
\bibitem [{\citenamefont {Hwang}\ \emph {et~al.}(2007)\citenamefont {Hwang},
  \citenamefont {Adam},\ and\ \citenamefont
  {Sarma}}]{das_sarma_carrier_transport_graphene}%
  \BibitemOpen
  \bibfield  {author} {\bibinfo {author} {\bibfnamefont {E.~H.}\ \bibnamefont
  {Hwang}}, \bibinfo {author} {\bibfnamefont {S.}~\bibnamefont {Adam}},\ and\
  \bibinfo {author} {\bibfnamefont {S.~D.}\ \bibnamefont {Sarma}},\ }\href
  {https://doi.org/10.1103/PhysRevLett.98.186806} {\bibfield  {journal}
  {\bibinfo  {journal} {Phys. Rev. Lett.}\ }\textbf {\bibinfo {volume} {98}},\
  \bibinfo {pages} {186806} (\bibinfo {year} {2007})}\BibitemShut {NoStop}%
\bibitem [{\citenamefont {Adam}\ \emph {et~al.}(2007)\citenamefont {Adam},
  \citenamefont {Hwang}, \citenamefont {Galitski},\ and\ \citenamefont
  {Das~Sarma}}]{das_sarma_graphene_transport_mean_field_theory}%
  \BibitemOpen
  \bibfield  {author} {\bibinfo {author} {\bibfnamefont {S.}~\bibnamefont
  {Adam}}, \bibinfo {author} {\bibfnamefont {E.~H.}\ \bibnamefont {Hwang}},
  \bibinfo {author} {\bibfnamefont {V.~M.}\ \bibnamefont {Galitski}},\ and\
  \bibinfo {author} {\bibfnamefont {S.}~\bibnamefont {Das~Sarma}},\ }\href
  {https://doi.org/10.1073/pnas.0704772104} {\bibfield  {journal} {\bibinfo
  {journal} {Proceedings of the National Academy of Sciences}\ }\textbf
  {\bibinfo {volume} {104}},\ \bibinfo {pages} {18392} (\bibinfo {year}
  {2007})}\BibitemShut {NoStop}%
\bibitem [{\citenamefont {Cheianov}\ \emph {et~al.}(2007)\citenamefont
  {Cheianov}, \citenamefont {Fal'ko}, \citenamefont {Altshuler},\ and\
  \citenamefont {Aleiner}}]{aleiner_graphene_conductivity}%
  \BibitemOpen
  \bibfield  {author} {\bibinfo {author} {\bibfnamefont {V.~V.}\ \bibnamefont
  {Cheianov}}, \bibinfo {author} {\bibfnamefont {V.~I.}\ \bibnamefont
  {Fal'ko}}, \bibinfo {author} {\bibfnamefont {B.~L.}\ \bibnamefont
  {Altshuler}},\ and\ \bibinfo {author} {\bibfnamefont {I.~L.}\ \bibnamefont
  {Aleiner}},\ }\href {https://doi.org/10.1103/PhysRevLett.99.176801}
  {\bibfield  {journal} {\bibinfo  {journal} {Phys. Rev. Lett.}\ }\textbf
  {\bibinfo {volume} {99}},\ \bibinfo {pages} {176801} (\bibinfo {year}
  {2007})}\BibitemShut {NoStop}%
\bibitem [{\citenamefont {Das~Sarma}\ \emph {et~al.}(2011)\citenamefont
  {Das~Sarma}, \citenamefont {Adam}, \citenamefont {Hwang},\ and\ \citenamefont
  {Rossi}}]{rossi_transport_graphene_review}%
  \BibitemOpen
  \bibfield  {author} {\bibinfo {author} {\bibfnamefont {S.}~\bibnamefont
  {Das~Sarma}}, \bibinfo {author} {\bibfnamefont {S.}~\bibnamefont {Adam}},
  \bibinfo {author} {\bibfnamefont {E.~H.}\ \bibnamefont {Hwang}},\ and\
  \bibinfo {author} {\bibfnamefont {E.}~\bibnamefont {Rossi}},\ }\href
  {https://doi.org/10.1103/RevModPhys.83.407} {\bibfield  {journal} {\bibinfo
  {journal} {Rev. Mod. Phys.}\ }\textbf {\bibinfo {volume} {83}},\ \bibinfo
  {pages} {407} (\bibinfo {year} {2011})}\BibitemShut {NoStop}%
\bibitem [{\citenamefont {Nersesyan}\ \emph {et~al.}(1995)\citenamefont
  {Nersesyan}, \citenamefont {Tsvelik},\ and\ \citenamefont
  {Wenger}}]{wenger_dos_disorder_2d_Dirac}%
  \BibitemOpen
  \bibfield  {author} {\bibinfo {author} {\bibfnamefont {A.}~\bibnamefont
  {Nersesyan}}, \bibinfo {author} {\bibfnamefont {A.}~\bibnamefont {Tsvelik}},\
  and\ \bibinfo {author} {\bibfnamefont {F.}~\bibnamefont {Wenger}},\ }\href
  {https://doi.org/https://doi.org/10.1016/0550-3213(95)00002-A} {\bibfield
  {journal} {\bibinfo  {journal} {Nuclear Physics B}\ }\textbf {\bibinfo
  {volume} {438}},\ \bibinfo {pages} {561} (\bibinfo {year}
  {1995})}\BibitemShut {NoStop}%
\bibitem [{\citenamefont {Nersesyan}\ \emph {et~al.}(1994)\citenamefont
  {Nersesyan}, \citenamefont {Tsvelik},\ and\ \citenamefont
  {Wenger}}]{wenger_disorder_defetcs_in_d-wave_SC}%
  \BibitemOpen
  \bibfield  {author} {\bibinfo {author} {\bibfnamefont {A.~A.}\ \bibnamefont
  {Nersesyan}}, \bibinfo {author} {\bibfnamefont {A.~M.}\ \bibnamefont
  {Tsvelik}},\ and\ \bibinfo {author} {\bibfnamefont {F.}~\bibnamefont
  {Wenger}},\ }\href {https://doi.org/10.1103/PhysRevLett.72.2628} {\bibfield
  {journal} {\bibinfo  {journal} {Phys. Rev. Lett.}\ }\textbf {\bibinfo
  {volume} {72}},\ \bibinfo {pages} {2628} (\bibinfo {year}
  {1994})}\BibitemShut {NoStop}%
\bibitem [{\citenamefont {Altland}\ \emph {et~al.}(2002)\citenamefont
  {Altland}, \citenamefont {Simons},\ and\ \citenamefont
  {Zirnbauer}}]{altland_d_wave_disorder}%
  \BibitemOpen
  \bibfield  {author} {\bibinfo {author} {\bibfnamefont {A.}~\bibnamefont
  {Altland}}, \bibinfo {author} {\bibfnamefont {B.}~\bibnamefont {Simons}},\
  and\ \bibinfo {author} {\bibfnamefont {M.}~\bibnamefont {Zirnbauer}},\ }\href
  {https://doi.org/https://doi.org/10.1016/S0370-1573(01)00065-5} {\bibfield
  {journal} {\bibinfo  {journal} {Physics Reports}\ }\textbf {\bibinfo {volume}
  {359}},\ \bibinfo {pages} {283} (\bibinfo {year} {2002})}\BibitemShut
  {NoStop}%
\bibitem [{\citenamefont {Novoselov}\ \emph {et~al.}(2004)\citenamefont
  {Novoselov}, \citenamefont {Geim}, \citenamefont {Morozov}, \citenamefont
  {Jiang}, \citenamefont {Zhang}, \citenamefont {Dubonos}, \citenamefont
  {Grigorieva},\ and\ \citenamefont {Firsov}}]{Novoselov+04}%
  \BibitemOpen
  \bibfield  {author} {\bibinfo {author} {\bibfnamefont {K.~S.}\ \bibnamefont
  {Novoselov}}, \bibinfo {author} {\bibfnamefont {A.~K.}\ \bibnamefont {Geim}},
  \bibinfo {author} {\bibfnamefont {S.~V.}\ \bibnamefont {Morozov}}, \bibinfo
  {author} {\bibfnamefont {D.}~\bibnamefont {Jiang}}, \bibinfo {author}
  {\bibfnamefont {Y.}~\bibnamefont {Zhang}}, \bibinfo {author} {\bibfnamefont
  {S.~V.}\ \bibnamefont {Dubonos}}, \bibinfo {author} {\bibfnamefont {I.~V.}\
  \bibnamefont {Grigorieva}},\ and\ \bibinfo {author} {\bibfnamefont {A.~A.}\
  \bibnamefont {Firsov}},\ }\href {https://doi.org/10.1126/science.1102896}
  {\bibfield  {journal} {\bibinfo  {journal} {Science}\ }\textbf {\bibinfo
  {volume} {306}},\ \bibinfo {pages} {666} (\bibinfo {year}
  {2004})}\BibitemShut {NoStop}%
\bibitem [{\citenamefont {Novoselov}\ \emph {et~al.}(2005)\citenamefont
  {Novoselov}, \citenamefont {Geim}, \citenamefont {Morozov}, \citenamefont
  {Jiang}, \citenamefont {Katsnelson}, \citenamefont {Grigorieva},
  \citenamefont {Dubonos},\ and\ \citenamefont {Firsov}}]{Novoselov+05}%
  \BibitemOpen
  \bibfield  {author} {\bibinfo {author} {\bibfnamefont {K.~S.}\ \bibnamefont
  {Novoselov}}, \bibinfo {author} {\bibfnamefont {A.~K.}\ \bibnamefont {Geim}},
  \bibinfo {author} {\bibfnamefont {S.~V.}\ \bibnamefont {Morozov}}, \bibinfo
  {author} {\bibfnamefont {D.}~\bibnamefont {Jiang}}, \bibinfo {author}
  {\bibfnamefont {M.~I.}\ \bibnamefont {Katsnelson}}, \bibinfo {author}
  {\bibfnamefont {I.~V.}\ \bibnamefont {Grigorieva}}, \bibinfo {author}
  {\bibfnamefont {S.~V.}\ \bibnamefont {Dubonos}},\ and\ \bibinfo {author}
  {\bibfnamefont {A.~A.}\ \bibnamefont {Firsov}},\ }\href
  {https://doi.org/10.1038/nature04233} {\bibfield  {journal} {\bibinfo
  {journal} {Nature}\ }\textbf {\bibinfo {volume} {438}},\ \bibinfo {pages}
  {197} (\bibinfo {year} {2005})}\BibitemShut {NoStop}%
\bibitem [{\citenamefont {Zhang}\ \emph {et~al.}(2005)\citenamefont {Zhang},
  \citenamefont {Tan}, \citenamefont {Stormer},\ and\ \citenamefont
  {Kim}}]{Zhang+05}%
  \BibitemOpen
  \bibfield  {author} {\bibinfo {author} {\bibfnamefont {Y.}~\bibnamefont
  {Zhang}}, \bibinfo {author} {\bibfnamefont {Y.-W.}\ \bibnamefont {Tan}},
  \bibinfo {author} {\bibfnamefont {H.~L.}\ \bibnamefont {Stormer}},\ and\
  \bibinfo {author} {\bibfnamefont {P.}~\bibnamefont {Kim}},\ }\href
  {https://doi.org/10.1038/nature04235} {\bibfield  {journal} {\bibinfo
  {journal} {Nature}\ }\textbf {\bibinfo {volume} {438}},\ \bibinfo {pages}
  {201} (\bibinfo {year} {2005})}\BibitemShut {NoStop}%
\bibitem [{\citenamefont {Ye}\ and\ \citenamefont
  {Sachdev}(1998)}]{sachdev_coulomb_quantum_hall}%
  \BibitemOpen
  \bibfield  {author} {\bibinfo {author} {\bibfnamefont {J.}~\bibnamefont
  {Ye}}\ and\ \bibinfo {author} {\bibfnamefont {S.}~\bibnamefont {Sachdev}},\
  }\href {https://doi.org/10.1103/PhysRevLett.80.5409} {\bibfield  {journal}
  {\bibinfo  {journal} {Phys. Rev. Lett.}\ }\textbf {\bibinfo {volume} {80}},\
  \bibinfo {pages} {5409} (\bibinfo {year} {1998})}\BibitemShut {NoStop}%
\bibitem [{\citenamefont {Ye}(1999)}]{ye_weak_disorder_quantum_hall}%
  \BibitemOpen
  \bibfield  {author} {\bibinfo {author} {\bibfnamefont {J.}~\bibnamefont
  {Ye}},\ }\href {https://doi.org/10.1103/PhysRevB.60.8290} {\bibfield
  {journal} {\bibinfo  {journal} {Phys. Rev. B}\ }\textbf {\bibinfo {volume}
  {60}},\ \bibinfo {pages} {8290} (\bibinfo {year} {1999})}\BibitemShut
  {NoStop}%
\bibitem [{\citenamefont {Stauber}\ \emph {et~al.}(2005)\citenamefont
  {Stauber}, \citenamefont {Guinea},\ and\ \citenamefont
  {Vozmediano}}]{vozmediano_disorder_graphene}%
  \BibitemOpen
  \bibfield  {author} {\bibinfo {author} {\bibfnamefont {T.}~\bibnamefont
  {Stauber}}, \bibinfo {author} {\bibfnamefont {F.}~\bibnamefont {Guinea}},\
  and\ \bibinfo {author} {\bibfnamefont {M.~A.~H.}\ \bibnamefont
  {Vozmediano}},\ }\href {https://doi.org/10.1103/PhysRevB.71.041406}
  {\bibfield  {journal} {\bibinfo  {journal} {Phys. Rev. B}\ }\textbf {\bibinfo
  {volume} {71}},\ \bibinfo {pages} {041406} (\bibinfo {year}
  {2005})}\BibitemShut {NoStop}%
\bibitem [{\citenamefont {Foster}\ and\ \citenamefont
  {Aleiner}(2008)}]{aleiner_graphene_large_N}%
  \BibitemOpen
  \bibfield  {author} {\bibinfo {author} {\bibfnamefont {M.~S.}\ \bibnamefont
  {Foster}}\ and\ \bibinfo {author} {\bibfnamefont {I.~L.}\ \bibnamefont
  {Aleiner}},\ }\href {https://doi.org/10.1103/PhysRevB.77.195413} {\bibfield
  {journal} {\bibinfo  {journal} {Phys. Rev. B}\ }\textbf {\bibinfo {volume}
  {77}},\ \bibinfo {pages} {195413} (\bibinfo {year} {2008})}\BibitemShut
  {NoStop}%
\bibitem [{\citenamefont {González}\ \emph {et~al.}(1993)\citenamefont
  {González}, \citenamefont {Guinea},\ and\ \citenamefont
  {Vozmediano}}]{Gonzalez+93}%
  \BibitemOpen
  \bibfield  {author} {\bibinfo {author} {\bibfnamefont {J.}~\bibnamefont
  {González}}, \bibinfo {author} {\bibfnamefont {F.}~\bibnamefont {Guinea}},\
  and\ \bibinfo {author} {\bibfnamefont {M.}~\bibnamefont {Vozmediano}},\
  }\href {https://doi.org/https://doi.org/10.1016/0550-3213(93)90009-E}
  {\bibfield  {journal} {\bibinfo  {journal} {Nuclear Physics B}\ }\textbf
  {\bibinfo {volume} {406}},\ \bibinfo {pages} {771} (\bibinfo {year}
  {1993})}\BibitemShut {NoStop}%
\bibitem [{\citenamefont {Lammert}\ and\ \citenamefont
  {Crespi}(2000)}]{Lammert+00}%
  \BibitemOpen
  \bibfield  {author} {\bibinfo {author} {\bibfnamefont {P.~E.}\ \bibnamefont
  {Lammert}}\ and\ \bibinfo {author} {\bibfnamefont {V.~H.}\ \bibnamefont
  {Crespi}},\ }\href {https://doi.org/10.1103/PhysRevLett.85.5190} {\bibfield
  {journal} {\bibinfo  {journal} {Phys. Rev. Lett.}\ }\textbf {\bibinfo
  {volume} {85}},\ \bibinfo {pages} {5190} (\bibinfo {year}
  {2000})}\BibitemShut {NoStop}%
\bibitem [{\citenamefont {Gonz\'alez}\ \emph {et~al.}(2001)\citenamefont
  {Gonz\'alez}, \citenamefont {Guinea},\ and\ \citenamefont
  {Vozmediano}}]{Gonzalez+01}%
  \BibitemOpen
  \bibfield  {author} {\bibinfo {author} {\bibfnamefont {J.}~\bibnamefont
  {Gonz\'alez}}, \bibinfo {author} {\bibfnamefont {F.}~\bibnamefont {Guinea}},\
  and\ \bibinfo {author} {\bibfnamefont {M.~A.~H.}\ \bibnamefont
  {Vozmediano}},\ }\href {https://doi.org/10.1103/PhysRevB.63.134421}
  {\bibfield  {journal} {\bibinfo  {journal} {Phys. Rev. B}\ }\textbf {\bibinfo
  {volume} {63}},\ \bibinfo {pages} {134421} (\bibinfo {year}
  {2001})}\BibitemShut {NoStop}%
\bibitem [{\citenamefont {Goswami}\ and\ \citenamefont
  {Chakravarty}(2011)}]{chakravarty_diffusive_metal}%
  \BibitemOpen
  \bibfield  {author} {\bibinfo {author} {\bibfnamefont {P.}~\bibnamefont
  {Goswami}}\ and\ \bibinfo {author} {\bibfnamefont {S.}~\bibnamefont
  {Chakravarty}},\ }\href {https://doi.org/10.1103/PhysRevLett.107.196803}
  {\bibfield  {journal} {\bibinfo  {journal} {Phys. Rev. Lett.}\ }\textbf
  {\bibinfo {volume} {107}},\ \bibinfo {pages} {196803} (\bibinfo {year}
  {2011})}\BibitemShut {NoStop}%
\bibitem [{\citenamefont
  {Gonz\'alez}(2017)}]{gonzalez_3d_weyl_coulomb_disorder}%
  \BibitemOpen
  \bibfield  {author} {\bibinfo {author} {\bibfnamefont {J.}~\bibnamefont
  {Gonz\'alez}},\ }\href {https://doi.org/10.1103/PhysRevB.96.081104}
  {\bibfield  {journal} {\bibinfo  {journal} {Phys. Rev. B}\ }\textbf {\bibinfo
  {volume} {96}},\ \bibinfo {pages} {081104} (\bibinfo {year}
  {2017})}\BibitemShut {NoStop}%
\bibitem [{\citenamefont {Roy}\ and\ \citenamefont
  {Das~Sarma}(2014)}]{das_sarma_disorder_3d_dirac}%
  \BibitemOpen
  \bibfield  {author} {\bibinfo {author} {\bibfnamefont {B.}~\bibnamefont
  {Roy}}\ and\ \bibinfo {author} {\bibfnamefont {S.}~\bibnamefont
  {Das~Sarma}},\ }\href {https://doi.org/10.1103/PhysRevB.90.241112} {\bibfield
   {journal} {\bibinfo  {journal} {Phys. Rev. B}\ }\textbf {\bibinfo {volume}
  {90}},\ \bibinfo {pages} {241112} (\bibinfo {year} {2014})}\BibitemShut
  {NoStop}%
\bibitem [{\citenamefont {Wang}\ \emph {et~al.}(2021)\citenamefont {Wang},
  \citenamefont {Li}, \citenamefont {Wang},\ and\ \citenamefont
  {Zhang}}]{wang_weyl_3d_coulomb_disorder}%
  \BibitemOpen
  \bibfield  {author} {\bibinfo {author} {\bibfnamefont {J.-R.}\ \bibnamefont
  {Wang}}, \bibinfo {author} {\bibfnamefont {W.}~\bibnamefont {Li}}, \bibinfo
  {author} {\bibfnamefont {G.}~\bibnamefont {Wang}},\ and\ \bibinfo {author}
  {\bibfnamefont {C.-J.}\ \bibnamefont {Zhang}},\ }\href
  {https://doi.org/10.1088/1361-648x/abd426} {\bibfield  {journal} {\bibinfo
  {journal} {Journal of Physics: Condensed Matter}\ }\textbf {\bibinfo {volume}
  {33}},\ \bibinfo {pages} {125601} (\bibinfo {year} {2021})}\BibitemShut
  {NoStop}%
\bibitem [{\citenamefont {Syzranov}\ and\ \citenamefont
  {Radzihovsky}(2018)}]{Syzranov+18}%
  \BibitemOpen
  \bibfield  {author} {\bibinfo {author} {\bibfnamefont {S.~V.}\ \bibnamefont
  {Syzranov}}\ and\ \bibinfo {author} {\bibfnamefont {L.}~\bibnamefont
  {Radzihovsky}},\ }\href
  {https://doi.org/10.1146/annurev-conmatphys-033117-054037} {\bibfield
  {journal} {\bibinfo  {journal} {Annual Review of Condensed Matter Physics}\
  }\textbf {\bibinfo {volume} {9}},\ \bibinfo {pages} {35} (\bibinfo {year}
  {2018})},\ \Eprint
  {https://arxiv.org/abs/https://doi.org/10.1146/annurev-conmatphys-033117-054037}
  {https://doi.org/10.1146/annurev-conmatphys-033117-054037} \BibitemShut
  {NoStop}%
\bibitem [{\citenamefont {Nandkishore}\ \emph {et~al.}(2014)\citenamefont
  {Nandkishore}, \citenamefont {Huse},\ and\ \citenamefont
  {Sondhi}}]{sondhi_rare_region_effects_dirac_3D}%
  \BibitemOpen
  \bibfield  {author} {\bibinfo {author} {\bibfnamefont {R.}~\bibnamefont
  {Nandkishore}}, \bibinfo {author} {\bibfnamefont {D.~A.}\ \bibnamefont
  {Huse}},\ and\ \bibinfo {author} {\bibfnamefont {S.~L.}\ \bibnamefont
  {Sondhi}},\ }\href {https://doi.org/10.1103/PhysRevB.89.245110} {\bibfield
  {journal} {\bibinfo  {journal} {Phys. Rev. B}\ }\textbf {\bibinfo {volume}
  {89}},\ \bibinfo {pages} {245110} (\bibinfo {year} {2014})}\BibitemShut
  {NoStop}%
\bibitem [{\citenamefont {Wilson}\ \emph {et~al.}(2020)\citenamefont {Wilson},
  \citenamefont {Huse}, \citenamefont {Das~Sarma},\ and\ \citenamefont
  {Pixley}}]{pixley_avoided_QCP_weyl}%
  \BibitemOpen
  \bibfield  {author} {\bibinfo {author} {\bibfnamefont {J.~H.}\ \bibnamefont
  {Wilson}}, \bibinfo {author} {\bibfnamefont {D.~A.}\ \bibnamefont {Huse}},
  \bibinfo {author} {\bibfnamefont {S.}~\bibnamefont {Das~Sarma}},\ and\
  \bibinfo {author} {\bibfnamefont {J.~H.}\ \bibnamefont {Pixley}},\ }\href
  {https://doi.org/10.1103/PhysRevB.102.100201} {\bibfield  {journal} {\bibinfo
   {journal} {Phys. Rev. B}\ }\textbf {\bibinfo {volume} {102}},\ \bibinfo
  {pages} {100201} (\bibinfo {year} {2020})}\BibitemShut {NoStop}%
\bibitem [{\citenamefont {Pixley}\ \emph {et~al.}(2016)\citenamefont {Pixley},
  \citenamefont {Huse},\ and\ \citenamefont {Das~Sarma}}]{Pixley+16}%
  \BibitemOpen
  \bibfield  {author} {\bibinfo {author} {\bibfnamefont {J.~H.}\ \bibnamefont
  {Pixley}}, \bibinfo {author} {\bibfnamefont {D.~A.}\ \bibnamefont {Huse}},\
  and\ \bibinfo {author} {\bibfnamefont {S.}~\bibnamefont {Das~Sarma}},\ }\href
  {https://doi.org/10.1103/PhysRevX.6.021042} {\bibfield  {journal} {\bibinfo
  {journal} {Phys. Rev. X}\ }\textbf {\bibinfo {volume} {6}},\ \bibinfo {pages}
  {021042} (\bibinfo {year} {2016})}\BibitemShut {NoStop}%
\bibitem [{\citenamefont {Buchhold}\ \emph
  {et~al.}(2018{\natexlab{a}})\citenamefont {Buchhold}, \citenamefont {Diehl},\
  and\ \citenamefont {Altland}}]{altland_dos_weyl_rare_region}%
  \BibitemOpen
  \bibfield  {author} {\bibinfo {author} {\bibfnamefont {M.}~\bibnamefont
  {Buchhold}}, \bibinfo {author} {\bibfnamefont {S.}~\bibnamefont {Diehl}},\
  and\ \bibinfo {author} {\bibfnamefont {A.}~\bibnamefont {Altland}},\ }\href
  {https://doi.org/10.1103/PhysRevLett.121.215301} {\bibfield  {journal}
  {\bibinfo  {journal} {Phys. Rev. Lett.}\ }\textbf {\bibinfo {volume} {121}},\
  \bibinfo {pages} {215301} (\bibinfo {year} {2018}{\natexlab{a}})}\BibitemShut
  {NoStop}%
\bibitem [{\citenamefont {Buchhold}\ \emph
  {et~al.}(2018{\natexlab{b}})\citenamefont {Buchhold}, \citenamefont {Diehl},\
  and\ \citenamefont {Altland}}]{altland_weyl_stable_against_disorder}%
  \BibitemOpen
  \bibfield  {author} {\bibinfo {author} {\bibfnamefont {M.}~\bibnamefont
  {Buchhold}}, \bibinfo {author} {\bibfnamefont {S.}~\bibnamefont {Diehl}},\
  and\ \bibinfo {author} {\bibfnamefont {A.}~\bibnamefont {Altland}},\ }\href
  {https://doi.org/10.1103/PhysRevB.98.205134} {\bibfield  {journal} {\bibinfo
  {journal} {Phys. Rev. B}\ }\textbf {\bibinfo {volume} {98}},\ \bibinfo
  {pages} {205134} (\bibinfo {year} {2018}{\natexlab{b}})}\BibitemShut
  {NoStop}%
\bibitem [{\citenamefont {Gross}\ and\ \citenamefont
  {Neveu}(1974)}]{gross_neveu_1974}%
  \BibitemOpen
  \bibfield  {author} {\bibinfo {author} {\bibfnamefont {D.~J.}\ \bibnamefont
  {Gross}}\ and\ \bibinfo {author} {\bibfnamefont {A.}~\bibnamefont {Neveu}},\
  }\href {https://doi.org/10.1103/PhysRevD.10.3235} {\bibfield  {journal}
  {\bibinfo  {journal} {Phys. Rev. D}\ }\textbf {\bibinfo {volume} {10}},\
  \bibinfo {pages} {3235} (\bibinfo {year} {1974})}\BibitemShut {NoStop}%
\bibitem [{\citenamefont
  {Zinn-Justin}(1991)}]{zinn-justin_four_fermion_interaction_1991}%
  \BibitemOpen
  \bibfield  {author} {\bibinfo {author} {\bibfnamefont {J.}~\bibnamefont
  {Zinn-Justin}},\ }\href
  {https://doi.org/https://doi.org/10.1016/0550-3213(91)90043-W} {\bibfield
  {journal} {\bibinfo  {journal} {Nuclear Physics B}\ }\textbf {\bibinfo
  {volume} {367}},\ \bibinfo {pages} {105 } (\bibinfo {year}
  {1991})}\BibitemShut {NoStop}%
\bibitem [{\citenamefont {Grushin}\ \emph {et~al.}(2013)\citenamefont
  {Grushin}, \citenamefont {Castro}, \citenamefont {Cortijo}, \citenamefont
  {de~Juan}, \citenamefont {Vozmediano},\ and\ \citenamefont
  {Valenzuela}}]{grushin_graphene_phases}%
  \BibitemOpen
  \bibfield  {author} {\bibinfo {author} {\bibfnamefont {A.~G.}\ \bibnamefont
  {Grushin}}, \bibinfo {author} {\bibfnamefont {E.~V.}\ \bibnamefont {Castro}},
  \bibinfo {author} {\bibfnamefont {A.}~\bibnamefont {Cortijo}}, \bibinfo
  {author} {\bibfnamefont {F.}~\bibnamefont {de~Juan}}, \bibinfo {author}
  {\bibfnamefont {M.~A.~H.}\ \bibnamefont {Vozmediano}},\ and\ \bibinfo
  {author} {\bibfnamefont {B.}~\bibnamefont {Valenzuela}},\ }\href
  {https://doi.org/10.1103/PhysRevB.87.085136} {\bibfield  {journal} {\bibinfo
  {journal} {Phys. Rev. B}\ }\textbf {\bibinfo {volume} {87}},\ \bibinfo
  {pages} {085136} (\bibinfo {year} {2013})}\BibitemShut {NoStop}%
\bibitem [{\citenamefont {Nandkishore}\ \emph {et~al.}(2013)\citenamefont
  {Nandkishore}, \citenamefont {Maciejko}, \citenamefont {Huse},\ and\
  \citenamefont {Sondhi}}]{sondhi_sc_of_disordered_Dirac}%
  \BibitemOpen
  \bibfield  {author} {\bibinfo {author} {\bibfnamefont {R.}~\bibnamefont
  {Nandkishore}}, \bibinfo {author} {\bibfnamefont {J.}~\bibnamefont
  {Maciejko}}, \bibinfo {author} {\bibfnamefont {D.~A.}\ \bibnamefont {Huse}},\
  and\ \bibinfo {author} {\bibfnamefont {S.~L.}\ \bibnamefont {Sondhi}},\
  }\href {https://doi.org/10.1103/PhysRevB.87.174511} {\bibfield  {journal}
  {\bibinfo  {journal} {Phys. Rev. B}\ }\textbf {\bibinfo {volume} {87}},\
  \bibinfo {pages} {174511} (\bibinfo {year} {2013})}\BibitemShut {NoStop}%
\bibitem [{\citenamefont {Yerzhakov}\ and\ \citenamefont
  {Maciejko}(2018)}]{maciejko_dirac_disorder_double_epsilon}%
  \BibitemOpen
  \bibfield  {author} {\bibinfo {author} {\bibfnamefont {H.}~\bibnamefont
  {Yerzhakov}}\ and\ \bibinfo {author} {\bibfnamefont {J.}~\bibnamefont
  {Maciejko}},\ }\href {https://doi.org/10.1103/PhysRevB.98.195142} {\bibfield
  {journal} {\bibinfo  {journal} {Phys. Rev. B}\ }\textbf {\bibinfo {volume}
  {98}},\ \bibinfo {pages} {195142} (\bibinfo {year} {2018})}\BibitemShut
  {NoStop}%
\bibitem [{\citenamefont {Yerzhakov}\ and\ \citenamefont
  {Maciejko}(2021)}]{maciejko_gny_random_mass_disorder}%
  \BibitemOpen
  \bibfield  {author} {\bibinfo {author} {\bibfnamefont {H.}~\bibnamefont
  {Yerzhakov}}\ and\ \bibinfo {author} {\bibfnamefont {J.}~\bibnamefont
  {Maciejko}},\ }\href
  {https://doi.org/https://doi.org/10.1016/j.nuclphysb.2020.115241} {\bibfield
  {journal} {\bibinfo  {journal} {Nuclear Physics B}\ }\textbf {\bibinfo
  {volume} {962}},\ \bibinfo {pages} {115241} (\bibinfo {year}
  {2021})}\BibitemShut {NoStop}%
\bibitem [{\citenamefont {Uryszek}\ \emph {et~al.}(2020)\citenamefont
  {Uryszek}, \citenamefont {Kr\"uger},\ and\ \citenamefont
  {Christou}}]{uryszek_soft_cut_off}%
  \BibitemOpen
  \bibfield  {author} {\bibinfo {author} {\bibfnamefont {M.~D.}\ \bibnamefont
  {Uryszek}}, \bibinfo {author} {\bibfnamefont {F.}~\bibnamefont {Kr\"uger}},\
  and\ \bibinfo {author} {\bibfnamefont {E.}~\bibnamefont {Christou}},\ }\href
  {https://doi.org/10.1103/PhysRevResearch.2.043265} {\bibfield  {journal}
  {\bibinfo  {journal} {Phys. Rev. Research}\ }\textbf {\bibinfo {volume}
  {2}},\ \bibinfo {pages} {043265} (\bibinfo {year} {2020})}\BibitemShut
  {NoStop}%
\bibitem [{\citenamefont {Chubukov}\ \emph {et~al.}(2004)\citenamefont
  {Chubukov}, \citenamefont {P\'epin},\ and\ \citenamefont
  {Rech}}]{chubukov_landau_damping}%
  \BibitemOpen
  \bibfield  {author} {\bibinfo {author} {\bibfnamefont {A.~V.}\ \bibnamefont
  {Chubukov}}, \bibinfo {author} {\bibfnamefont {C.}~\bibnamefont {P\'epin}},\
  and\ \bibinfo {author} {\bibfnamefont {J.}~\bibnamefont {Rech}},\ }\href
  {https://doi.org/10.1103/PhysRevLett.92.147003} {\bibfield  {journal}
  {\bibinfo  {journal} {Phys. Rev. Lett.}\ }\textbf {\bibinfo {volume} {92}},\
  \bibinfo {pages} {147003} (\bibinfo {year} {2004})}\BibitemShut {NoStop}%
\bibitem [{\citenamefont {Edwards}\ and\ \citenamefont
  {Anderson}(1975)}]{edwards_anderson_spin_glasses_1975}%
  \BibitemOpen
  \bibfield  {author} {\bibinfo {author} {\bibfnamefont {S.~F.}\ \bibnamefont
  {Edwards}}\ and\ \bibinfo {author} {\bibfnamefont {P.~W.}\ \bibnamefont
  {Anderson}},\ }\href {https://doi.org/10.1088/0305-4608/5/5/017} {\bibfield
  {journal} {\bibinfo  {journal} {Journal of Physics F: Metal Physics}\
  }\textbf {\bibinfo {volume} {5}},\ \bibinfo {pages} {965} (\bibinfo {year}
  {1975})}\BibitemShut {NoStop}%
\bibitem [{\citenamefont {Fischer}\ and\ \citenamefont
  {Hertz}(1991)}]{fischer_hertz_sping_glasses}%
  \BibitemOpen
  \bibfield  {author} {\bibinfo {author} {\bibfnamefont {K.~H.}\ \bibnamefont
  {Fischer}}\ and\ \bibinfo {author} {\bibfnamefont {J.~A.}\ \bibnamefont
  {Hertz}},\ }\href {https://doi.org/10.1017/CBO9780511628771} {\emph {\bibinfo
  {title} {Spin Glasses}}},\ Cambridge Studies in Magnetism\ (\bibinfo
  {publisher} {Cambridge University Press},\ \bibinfo {year}
  {1991})\BibitemShut {NoStop}%
\bibitem [{\citenamefont {Lee}(2009)}]{Lee+09}%
  \BibitemOpen
  \bibfield  {author} {\bibinfo {author} {\bibfnamefont {S.-S.}\ \bibnamefont
  {Lee}},\ }\href {https://doi.org/10.1103/PhysRevB.80.165102} {\bibfield
  {journal} {\bibinfo  {journal} {Phys. Rev. B}\ }\textbf {\bibinfo {volume}
  {80}},\ \bibinfo {pages} {165102} (\bibinfo {year} {2009})}\BibitemShut
  {NoStop}%
\bibitem [{\citenamefont {Herbut}\ \emph {et~al.}(2008)\citenamefont {Herbut},
  \citenamefont {Juricic},\ and\ \citenamefont
  {Vafek}}]{vafek_coulomb_ripples_graphene}%
  \BibitemOpen
  \bibfield  {author} {\bibinfo {author} {\bibfnamefont {I.~F.}\ \bibnamefont
  {Herbut}}, \bibinfo {author} {\bibfnamefont {V.}~\bibnamefont {Juricic}},\
  and\ \bibinfo {author} {\bibfnamefont {O.}~\bibnamefont {Vafek}},\ }\href
  {https://doi.org/10.1103/PhysRevLett.100.046403} {\bibfield  {journal}
  {\bibinfo  {journal} {Phys. Rev. Lett.}\ }\textbf {\bibinfo {volume} {100}},\
  \bibinfo {pages} {046403} (\bibinfo {year} {2008})}\BibitemShut {NoStop}%
\bibitem [{\citenamefont {Vafek}\ and\ \citenamefont
  {Case}(2008)}]{case_2d_Dirac_Coulomb_random_gauge_potential}%
  \BibitemOpen
  \bibfield  {author} {\bibinfo {author} {\bibfnamefont {O.}~\bibnamefont
  {Vafek}}\ and\ \bibinfo {author} {\bibfnamefont {M.~J.}\ \bibnamefont
  {Case}},\ }\href {https://doi.org/10.1103/PhysRevB.77.033410} {\bibfield
  {journal} {\bibinfo  {journal} {Phys. Rev. B}\ }\textbf {\bibinfo {volume}
  {77}},\ \bibinfo {pages} {033410} (\bibinfo {year} {2008})}\BibitemShut
  {NoStop}%
\bibitem [{\citenamefont {Ludwig}\ \emph {et~al.}(1994)\citenamefont {Ludwig},
  \citenamefont {Fisher}, \citenamefont {Shankar},\ and\ \citenamefont
  {Grinstein}}]{grinstein_qhe_disorder}%
  \BibitemOpen
  \bibfield  {author} {\bibinfo {author} {\bibfnamefont {A.~W.~W.}\
  \bibnamefont {Ludwig}}, \bibinfo {author} {\bibfnamefont {M.~P.~A.}\
  \bibnamefont {Fisher}}, \bibinfo {author} {\bibfnamefont {R.}~\bibnamefont
  {Shankar}},\ and\ \bibinfo {author} {\bibfnamefont {G.}~\bibnamefont
  {Grinstein}},\ }\href {https://doi.org/10.1103/PhysRevB.50.7526} {\bibfield
  {journal} {\bibinfo  {journal} {Phys. Rev. B}\ }\textbf {\bibinfo {volume}
  {50}},\ \bibinfo {pages} {7526} (\bibinfo {year} {1994})}\BibitemShut
  {NoStop}%
\bibitem [{\citenamefont {Vasil'ev}\ and\ \citenamefont
  {Stepanenko}(1993)}]{Vasilev_nu_n2_1993}%
  \BibitemOpen
  \bibfield  {author} {\bibinfo {author} {\bibfnamefont {A.~N.}\ \bibnamefont
  {Vasil'ev}}\ and\ \bibinfo {author} {\bibfnamefont {A.}~\bibnamefont
  {Stepanenko}},\ }\href@noop {} {\bibfield  {journal} {\bibinfo  {journal}
  {Theoretical and Mathematical Physics}\ }\textbf {\bibinfo {volume} {97}},\
  \bibinfo {pages} {1349} (\bibinfo {year} {1993})}\BibitemShut {NoStop}%
\bibitem [{\citenamefont {Gracey}(1994)}]{Gracey94}%
  \BibitemOpen
  \bibfield  {author} {\bibinfo {author} {\bibfnamefont {J.}~\bibnamefont
  {Gracey}},\ }\href {https://doi.org/10.1142/S0217751X94000340} {\bibfield
  {journal} {\bibinfo  {journal} {International Journal of Modern Physics A}\
  }\textbf {\bibinfo {volume} {9}},\ \bibinfo {pages} {727} (\bibinfo {year}
  {1994})}\BibitemShut {NoStop}%
\bibitem [{\citenamefont {Iliesiu}\ \emph {et~al.}(2018)\citenamefont
  {Iliesiu}, \citenamefont {Kos}, \citenamefont {Poland}, \citenamefont
  {Pufu},\ and\ \citenamefont {Simmons-Duffin}}]{Iliesiu+18}%
  \BibitemOpen
  \bibfield  {author} {\bibinfo {author} {\bibfnamefont {L.}~\bibnamefont
  {Iliesiu}}, \bibinfo {author} {\bibfnamefont {F.}~\bibnamefont {Kos}},
  \bibinfo {author} {\bibfnamefont {D.}~\bibnamefont {Poland}}, \bibinfo
  {author} {\bibfnamefont {S.~S.}\ \bibnamefont {Pufu}},\ and\ \bibinfo
  {author} {\bibfnamefont {D.}~\bibnamefont {Simmons-Duffin}},\ }\href
  {https://doi.org/10.1007/JHEP01(2018)036} {\bibfield  {journal} {\bibinfo
  {journal} {Journal of High Energy Physics}\ }\textbf {\bibinfo {volume}
  {2018}},\ \bibinfo {pages} {36} (\bibinfo {year} {2018})}\BibitemShut
  {NoStop}%
\bibitem [{\citenamefont {Gat}\ \emph {et~al.}(1990)\citenamefont {Gat},
  \citenamefont {Kovner}, \citenamefont {Rosenstein},\ and\ \citenamefont
  {Warr}}]{Gat1990}%
  \BibitemOpen
  \bibfield  {author} {\bibinfo {author} {\bibfnamefont {G.}~\bibnamefont
  {Gat}}, \bibinfo {author} {\bibfnamefont {A.}~\bibnamefont {Kovner}},
  \bibinfo {author} {\bibfnamefont {B.}~\bibnamefont {Rosenstein}},\ and\
  \bibinfo {author} {\bibfnamefont {B.}~\bibnamefont {Warr}},\ }\href
  {https://doi.org/https://doi.org/10.1016/0370-2693(90)90425-6} {\bibfield
  {journal} {\bibinfo  {journal} {Physics Letters B}\ }\textbf {\bibinfo
  {volume} {240}},\ \bibinfo {pages} {158 } (\bibinfo {year}
  {1990})}\BibitemShut {NoStop}%
\bibitem [{\citenamefont {Gracey}(1991)}]{Gracey_eta_n2_1991}%
  \BibitemOpen
  \bibfield  {author} {\bibinfo {author} {\bibfnamefont {J.}~\bibnamefont
  {Gracey}},\ }\href {https://doi.org/10.1142/S0217751X91000241} {\bibfield
  {journal} {\bibinfo  {journal} {International Journal of Modern Physics A}\
  }\textbf {\bibinfo {volume} {06}},\ \bibinfo {pages} {395} (\bibinfo {year}
  {1991})}\BibitemShut {NoStop}%
\bibitem [{\citenamefont {Gracey}(1992)}]{Gracey_nu_n2_1992}%
  \BibitemOpen
  \bibfield  {author} {\bibinfo {author} {\bibfnamefont {J.}~\bibnamefont
  {Gracey}},\ }\href
  {https://doi.org/https://doi.org/10.1016/0370-2693(92)91265-B} {\bibfield
  {journal} {\bibinfo  {journal} {Physics Letters B}\ }\textbf {\bibinfo
  {volume} {297}},\ \bibinfo {pages} {293 } (\bibinfo {year}
  {1992})}\BibitemShut {NoStop}%
\bibitem [{\citenamefont {Harris}(1974)}]{Harris74}%
  \BibitemOpen
  \bibfield  {author} {\bibinfo {author} {\bibfnamefont {A.~B.}\ \bibnamefont
  {Harris}},\ }\href {https://doi.org/10.1088/0022-3719/7/9/009} {\bibfield
  {journal} {\bibinfo  {journal} {Journal of Physics C: Solid State Physics}\
  }\textbf {\bibinfo {volume} {7}},\ \bibinfo {pages} {1671} (\bibinfo {year}
  {1974})}\BibitemShut {NoStop}%
\bibitem [{\citenamefont {Chayes}\ \emph {et~al.}(1986)\citenamefont {Chayes},
  \citenamefont {Chayes}, \citenamefont {Fisher},\ and\ \citenamefont
  {Spencer}}]{Chayes+86}%
  \BibitemOpen
  \bibfield  {author} {\bibinfo {author} {\bibfnamefont {J.~T.}\ \bibnamefont
  {Chayes}}, \bibinfo {author} {\bibfnamefont {L.}~\bibnamefont {Chayes}},
  \bibinfo {author} {\bibfnamefont {D.~S.}\ \bibnamefont {Fisher}},\ and\
  \bibinfo {author} {\bibfnamefont {T.}~\bibnamefont {Spencer}},\ }\href
  {https://doi.org/10.1103/PhysRevLett.57.2999} {\bibfield  {journal} {\bibinfo
   {journal} {Phys. Rev. Lett.}\ }\textbf {\bibinfo {volume} {57}},\ \bibinfo
  {pages} {2999} (\bibinfo {year} {1986})}\BibitemShut {NoStop}%
\bibitem [{\citenamefont {Herbut}\ \emph {et~al.}(2009)\citenamefont {Herbut},
  \citenamefont {Juriicic},\ and\ \citenamefont {Roy}}]{Herbut+09}%
  \BibitemOpen
  \bibfield  {author} {\bibinfo {author} {\bibfnamefont {I.~F.}\ \bibnamefont
  {Herbut}}, \bibinfo {author} {\bibfnamefont {V.}~\bibnamefont {Juriicic}},\
  and\ \bibinfo {author} {\bibfnamefont {B.}~\bibnamefont {Roy}},\ }\href
  {https://doi.org/10.1103/PhysRevB.79.085116} {\bibfield  {journal} {\bibinfo
  {journal} {Phys. Rev. B}\ }\textbf {\bibinfo {volume} {79}},\ \bibinfo
  {pages} {085116} (\bibinfo {year} {2009})}\BibitemShut {NoStop}%
\bibitem [{\citenamefont {Imry}\ and\ \citenamefont {Ma}(1975)}]{Imry+75}%
  \BibitemOpen
  \bibfield  {author} {\bibinfo {author} {\bibfnamefont {Y.}~\bibnamefont
  {Imry}}\ and\ \bibinfo {author} {\bibfnamefont {S.-k.}\ \bibnamefont {Ma}},\
  }\href {https://doi.org/10.1103/PhysRevLett.35.1399} {\bibfield  {journal}
  {\bibinfo  {journal} {Phys. Rev. Lett.}\ }\textbf {\bibinfo {volume} {35}},\
  \bibinfo {pages} {1399} (\bibinfo {year} {1975})}\BibitemShut {NoStop}%
\bibitem [{\citenamefont {Aharony}\ \emph {et~al.}(1976)\citenamefont
  {Aharony}, \citenamefont {Imry},\ and\ \citenamefont {Ma}}]{Aharony+76}%
  \BibitemOpen
  \bibfield  {author} {\bibinfo {author} {\bibfnamefont {A.}~\bibnamefont
  {Aharony}}, \bibinfo {author} {\bibfnamefont {Y.}~\bibnamefont {Imry}},\ and\
  \bibinfo {author} {\bibfnamefont {S.-k.}\ \bibnamefont {Ma}},\ }\href
  {https://doi.org/10.1103/PhysRevLett.37.1364} {\bibfield  {journal} {\bibinfo
   {journal} {Phys. Rev. Lett.}\ }\textbf {\bibinfo {volume} {37}},\ \bibinfo
  {pages} {1364} (\bibinfo {year} {1976})}\BibitemShut {NoStop}%
\bibitem [{\citenamefont {Aizenman}\ and\ \citenamefont
  {Wehr}(1989)}]{Aizenman+89}%
  \BibitemOpen
  \bibfield  {author} {\bibinfo {author} {\bibfnamefont {M.}~\bibnamefont
  {Aizenman}}\ and\ \bibinfo {author} {\bibfnamefont {J.}~\bibnamefont
  {Wehr}},\ }\href {https://doi.org/10.1103/PhysRevLett.62.2503} {\bibfield
  {journal} {\bibinfo  {journal} {Phys. Rev. Lett.}\ }\textbf {\bibinfo
  {volume} {62}},\ \bibinfo {pages} {2503} (\bibinfo {year}
  {1989})}\BibitemShut {NoStop}%
\bibitem [{\citenamefont {Parisi}(1979)}]{parisi_replica_order_parameter_1979}%
  \BibitemOpen
  \bibfield  {author} {\bibinfo {author} {\bibfnamefont {G.}~\bibnamefont
  {Parisi}},\ }\href {https://doi.org/10.1103/PhysRevLett.43.1754} {\bibfield
  {journal} {\bibinfo  {journal} {Phys. Rev. Lett.}\ }\textbf {\bibinfo
  {volume} {43}},\ \bibinfo {pages} {1754} (\bibinfo {year}
  {1979})}\BibitemShut {NoStop}%
\bibitem [{\citenamefont {Pixley}\ and\ \citenamefont
  {Wilson}(2021)}]{pixley_rare_region_effects_review}%
  \BibitemOpen
  \bibfield  {author} {\bibinfo {author} {\bibfnamefont {J.}~\bibnamefont
  {Pixley}}\ and\ \bibinfo {author} {\bibfnamefont {J.~H.}\ \bibnamefont
  {Wilson}},\ }\href
  {https://doi.org/https://doi.org/10.1016/j.aop.2021.168455} {\bibfield
  {journal} {\bibinfo  {journal} {Annals of Physics}\ ,\ \bibinfo {pages}
  {168455}} (\bibinfo {year} {2021})}\BibitemShut {NoStop}%
\bibitem [{\citenamefont {Zhao}\ \emph {et~al.}(2017)\citenamefont {Zhao},
  \citenamefont {Wang},\ and\ \citenamefont {Liu}}]{Zhao+17}%
  \BibitemOpen
  \bibfield  {author} {\bibinfo {author} {\bibfnamefont {P.-L.}\ \bibnamefont
  {Zhao}}, \bibinfo {author} {\bibfnamefont {A.-M.}\ \bibnamefont {Wang}},\
  and\ \bibinfo {author} {\bibfnamefont {G.-Z.}\ \bibnamefont {Liu}},\ }\href
  {https://doi.org/10.1103/PhysRevB.95.235144} {\bibfield  {journal} {\bibinfo
  {journal} {Phys. Rev. B}\ }\textbf {\bibinfo {volume} {95}},\ \bibinfo
  {pages} {235144} (\bibinfo {year} {2017})}\BibitemShut {NoStop}%
\bibitem [{\citenamefont {Goswami}\ \emph {et~al.}(2017)\citenamefont
  {Goswami}, \citenamefont {Goldman},\ and\ \citenamefont
  {Raghu}}]{Goswami+17}%
  \BibitemOpen
  \bibfield  {author} {\bibinfo {author} {\bibfnamefont {P.}~\bibnamefont
  {Goswami}}, \bibinfo {author} {\bibfnamefont {H.}~\bibnamefont {Goldman}},\
  and\ \bibinfo {author} {\bibfnamefont {S.}~\bibnamefont {Raghu}},\ }\href
  {https://doi.org/10.1103/PhysRevB.95.235145} {\bibfield  {journal} {\bibinfo
  {journal} {Phys. Rev. B}\ }\textbf {\bibinfo {volume} {95}},\ \bibinfo
  {pages} {235145} (\bibinfo {year} {2017})}\BibitemShut {NoStop}%
\bibitem [{\citenamefont {Thomson}\ and\ \citenamefont
  {Sachdev}(2017)}]{Thomson+17}%
  \BibitemOpen
  \bibfield  {author} {\bibinfo {author} {\bibfnamefont {A.}~\bibnamefont
  {Thomson}}\ and\ \bibinfo {author} {\bibfnamefont {S.}~\bibnamefont
  {Sachdev}},\ }\href {https://doi.org/10.1103/PhysRevB.95.235146} {\bibfield
  {journal} {\bibinfo  {journal} {Phys. Rev. B}\ }\textbf {\bibinfo {volume}
  {95}},\ \bibinfo {pages} {235146} (\bibinfo {year} {2017})}\BibitemShut
  {NoStop}%
\bibitem [{\citenamefont {Appelquist}\ \emph {et~al.}(1986)\citenamefont
  {Appelquist}, \citenamefont {Bowick}, \citenamefont {Karabali},\ and\
  \citenamefont {Wijewardhana}}]{Appelquist+86}%
  \BibitemOpen
  \bibfield  {author} {\bibinfo {author} {\bibfnamefont {T.~W.}\ \bibnamefont
  {Appelquist}}, \bibinfo {author} {\bibfnamefont {M.}~\bibnamefont {Bowick}},
  \bibinfo {author} {\bibfnamefont {D.}~\bibnamefont {Karabali}},\ and\
  \bibinfo {author} {\bibfnamefont {L.~C.~R.}\ \bibnamefont {Wijewardhana}},\
  }\href {https://doi.org/10.1103/PhysRevD.33.3704} {\bibfield  {journal}
  {\bibinfo  {journal} {Phys. Rev. D}\ }\textbf {\bibinfo {volume} {33}},\
  \bibinfo {pages} {3704} (\bibinfo {year} {1986})}\BibitemShut {NoStop}%
\bibitem [{\citenamefont {Appelquist}\ \emph {et~al.}(1988)\citenamefont
  {Appelquist}, \citenamefont {Nash},\ and\ \citenamefont
  {Wijewardhana}}]{Appelquist+88}%
  \BibitemOpen
  \bibfield  {author} {\bibinfo {author} {\bibfnamefont {T.}~\bibnamefont
  {Appelquist}}, \bibinfo {author} {\bibfnamefont {D.}~\bibnamefont {Nash}},\
  and\ \bibinfo {author} {\bibfnamefont {L.~C.~R.}\ \bibnamefont
  {Wijewardhana}},\ }\href {https://doi.org/10.1103/PhysRevLett.60.2575}
  {\bibfield  {journal} {\bibinfo  {journal} {Phys. Rev. Lett.}\ }\textbf
  {\bibinfo {volume} {60}},\ \bibinfo {pages} {2575} (\bibinfo {year}
  {1988})}\BibitemShut {NoStop}%
\bibitem [{\citenamefont {Nash}(1989)}]{Nash89}%
  \BibitemOpen
  \bibfield  {author} {\bibinfo {author} {\bibfnamefont {D.}~\bibnamefont
  {Nash}},\ }\href {https://doi.org/10.1103/PhysRevLett.62.3024} {\bibfield
  {journal} {\bibinfo  {journal} {Phys. Rev. Lett.}\ }\textbf {\bibinfo
  {volume} {62}},\ \bibinfo {pages} {3024} (\bibinfo {year}
  {1989})}\BibitemShut {NoStop}%
\bibitem [{\citenamefont {Maris}(1996)}]{Maris96}%
  \BibitemOpen
  \bibfield  {author} {\bibinfo {author} {\bibfnamefont {P.}~\bibnamefont
  {Maris}},\ }\href {https://doi.org/10.1103/PhysRevD.54.4049} {\bibfield
  {journal} {\bibinfo  {journal} {Phys. Rev. D}\ }\textbf {\bibinfo {volume}
  {54}},\ \bibinfo {pages} {4049} (\bibinfo {year} {1996})}\BibitemShut
  {NoStop}%
\bibitem [{\citenamefont {Goldman}\ \emph {et~al.}(2020)\citenamefont
  {Goldman}, \citenamefont {Thomson}, \citenamefont {Nie},\ and\ \citenamefont
  {Bi}}]{Goldman+20}%
  \BibitemOpen
  \bibfield  {author} {\bibinfo {author} {\bibfnamefont {H.}~\bibnamefont
  {Goldman}}, \bibinfo {author} {\bibfnamefont {A.}~\bibnamefont {Thomson}},
  \bibinfo {author} {\bibfnamefont {L.}~\bibnamefont {Nie}},\ and\ \bibinfo
  {author} {\bibfnamefont {Z.}~\bibnamefont {Bi}},\ }\href
  {https://doi.org/10.1103/PhysRevB.101.144506} {\bibfield  {journal} {\bibinfo
   {journal} {Phys. Rev. B}\ }\textbf {\bibinfo {volume} {101}},\ \bibinfo
  {pages} {144506} (\bibinfo {year} {2020})}\BibitemShut {NoStop}%
\bibitem [{\citenamefont {Martin}\ \emph {et~al.}(2008)\citenamefont {Martin},
  \citenamefont {Akerman}, \citenamefont {Ulbricht}, \citenamefont {Lohmann},
  \citenamefont {Smet}, \citenamefont {von Klitzing},\ and\ \citenamefont
  {Yacoby}}]{martin_puddles_graphene_stm}%
  \BibitemOpen
  \bibfield  {author} {\bibinfo {author} {\bibfnamefont {J.}~\bibnamefont
  {Martin}}, \bibinfo {author} {\bibfnamefont {N.}~\bibnamefont {Akerman}},
  \bibinfo {author} {\bibfnamefont {G.}~\bibnamefont {Ulbricht}}, \bibinfo
  {author} {\bibfnamefont {T.}~\bibnamefont {Lohmann}}, \bibinfo {author}
  {\bibfnamefont {J.~H.}\ \bibnamefont {Smet}}, \bibinfo {author}
  {\bibfnamefont {K.}~\bibnamefont {von Klitzing}},\ and\ \bibinfo {author}
  {\bibfnamefont {A.}~\bibnamefont {Yacoby}},\ }\href
  {https://doi.org/10.1038/nphys781} {\bibfield  {journal} {\bibinfo  {journal}
  {Nature Physics}\ }\textbf {\bibinfo {volume} {4}},\ \bibinfo {pages} {144}
  (\bibinfo {year} {2008})}\BibitemShut {NoStop}%
\bibitem [{\citenamefont {Deshpande}\ \emph {et~al.}(2009)\citenamefont
  {Deshpande}, \citenamefont {Bao}, \citenamefont {Miao}, \citenamefont {Lau},\
  and\ \citenamefont {LeRoy}}]{deshpande_spectroscopyt}%
  \BibitemOpen
  \bibfield  {author} {\bibinfo {author} {\bibfnamefont {A.}~\bibnamefont
  {Deshpande}}, \bibinfo {author} {\bibfnamefont {W.}~\bibnamefont {Bao}},
  \bibinfo {author} {\bibfnamefont {F.}~\bibnamefont {Miao}}, \bibinfo {author}
  {\bibfnamefont {C.~N.}\ \bibnamefont {Lau}},\ and\ \bibinfo {author}
  {\bibfnamefont {B.~J.}\ \bibnamefont {LeRoy}},\ }\href
  {https://doi.org/10.1103/PhysRevB.79.205411} {\bibfield  {journal} {\bibinfo
  {journal} {Phys. Rev. B}\ }\textbf {\bibinfo {volume} {79}},\ \bibinfo
  {pages} {205411} (\bibinfo {year} {2009})}\BibitemShut {NoStop}%
\bibitem [{\citenamefont {Zhang}\ \emph {et~al.}(2009)\citenamefont {Zhang},
  \citenamefont {Brar}, \citenamefont {Girit}, \citenamefont {Zettl},\ and\
  \citenamefont {Crommie}}]{zhang_graphene_inhomogeneity}%
  \BibitemOpen
  \bibfield  {author} {\bibinfo {author} {\bibfnamefont {Y.}~\bibnamefont
  {Zhang}}, \bibinfo {author} {\bibfnamefont {V.~W.}\ \bibnamefont {Brar}},
  \bibinfo {author} {\bibfnamefont {C.}~\bibnamefont {Girit}}, \bibinfo
  {author} {\bibfnamefont {A.}~\bibnamefont {Zettl}},\ and\ \bibinfo {author}
  {\bibfnamefont {M.~F.}\ \bibnamefont {Crommie}},\ }\href
  {https://doi.org/10.1038/nphys1365} {\bibfield  {journal} {\bibinfo
  {journal} {Nature Physics}\ }\textbf {\bibinfo {volume} {5}},\ \bibinfo
  {pages} {722} (\bibinfo {year} {2009})}\BibitemShut {NoStop}%
\end{thebibliography}
\end{document}